\documentclass[preprint,11pt]{elsarticle}

\usepackage{amssymb}
\usepackage{array}
\usepackage{comment}
\usepackage{textcomp}
\usepackage{mathtools,multirow,amssymb,commath,siunitx,gensymb} 
\usepackage{amsmath}

\journal{Journal of the European Ceramic Society}

\begin{document}

\begin{frontmatter}



\title{High- and medium-entropy nitride coatings from the Cr-Hf-Mo-Ta-W-N system: properties and high-temperature stability}


\author[inst1]{Pavel Souček}
\author[inst1]{Stanislava Debnárová}
\author[inst2]{Šárka Zuzjaková}
\author[inst3,inst5]{Shuyao Lin}
\author[inst1]{Matej Fekete}
\author[inst4]{Zsolt Czigány}
\author[inst4]{Katalin Balázsi}
\author[inst1]{Lukáš Vrána}
\author[inst1]{Tatiana Pitoňáková}
\author[inst1]{Ondřej Jašek}
\author[inst2]{Petr Zeman}
\author[inst3,inst5]{Nikola Koutná}

\affiliation[inst1]{organization={Department of Plasma Physics and Technology, Faculty of Science, Masaryk University},
            addressline={Kotlářská 2}, 
            city={Brno},
            postcode={61137}, 
            country={Czech Republic}}

\affiliation[inst2]{organization={Department of Physics and NTIS – European Centre of Excellence, University of West Bohemia in Pilsen},
            addressline={Univerzitní 8}, 
            city={Pilsen},
            postcode={30100}, 
            country={Czech Republic}}

\affiliation[inst3]{organization={Institute of Materials Science and Technology, Faculty of Mechanical and Industrial Engineering, Technische Universität Wien},
            addressline={Getreidemarkt 9}, 
            city={Vienna},
            postcode={A-1060}, 
            country={Austria}}

\affiliation[inst5]{organization={Linköping University, Department of Physics, Chemistry, and Biology (IFM)},
            addressline={Olaus Magnus väg}, 
            city={Linköping},
            postcode={SE-58183}, 
            country={Sweden}}

\affiliation[inst4]{organization={Institute of Technical Physics and Materials Science, HUN-REN Centre for Energy Research},
            addressline={Konkoly Thege M. út 29-33}, 
            city={Budapest},
            postcode={H-1121}, 
            country={Hungary}}

\begin{abstract}
High- and medium-entropy nitride coatings from the Cr–Hf–Mo–Ta–W–N system were studied using \textit{ab initio} calculations and experiments to clarify the role of entropy and individual elements in phase stability, microstructure, and high-temperature behaviour. Formation energy calculations indicated that nitrogen vacancies stabilise the cubic (fcc) phase, with hafnium and tantalum acting as strong stabilisers, while tungsten destabilises the lattice.  
Coatings were deposited by reactive magnetron sputtering at $\sim$50\textdegree C (AT) and $\sim$580\textdegree C (HT). All exhibited columnar fcc structures; high-temperature deposition produced denser coatings, lower nitrogen content, and larger crystallites, resulting in higher hardness and elastic modulus.  
Thermal stability was tested up to 1200\textdegree C on Si and oxidation at 1400\textdegree C on sapphire. AT coatings failed early, while most HT coatings endured. Nitrogen loss $\lesssim$10~at.\% at 1000\textdegree C was critical for survival. TEM revealed tungsten segregation and HfO$_2$ formation, while fcc nitride remained dominant. Ta enrichment proved essential for superior thermal and oxidation stability.

\end{abstract}


\begin{highlights}
\item {\it{Ab initio}} calculations indicate stabilizing (destabilizing) effect of Hf and Ta (W).
\item Properties of the as-deposited coatings are governed by their microstructure.
\item Nitrogen release during annealing leads to a high concentration of vacancies.
\item Tungsten segregates and hafnium forms oxides during oxidation.
\item Not entropy, but tantalum content is critical for achieving the best stability.

\end{highlights}


\end{frontmatter}


\section{Introduction}
Since the Bronze Age, new alloys have been created by alloying a single principal element material with a small amount of other elements. This, by necessity, limits the number of achievable alloys by the number of usable principal elements. The properties of the resulting alloys are heavily dependent on the properties of the principal element, and alloying elements are used to promote its desirable properties or mitigate the undesirable ones. In 2004, a novel paradigm for alloy design was introduced, which left behind the reliance on a single principal element in favour of using multiple principal elements in near-equiatomic ratios \cite{CANTOR2004213,yeh04}. These so-called baseless or multi-principal element alloys increase the entropic contribution to the Gibbs free energy, allowing the formation of stable, chemically complex materials as simple solid solutions—typically with cubic or hexagonal crystal structures \cite{MIRACLE2017448}. This stabilisation effect generally arises when five or more elements are combined. Such materials are known as high entropy alloys (HEAs). Their behaviour is governed by four fundamental effects: the high entropy effect, sluggish diffusion, lattice distortion, and the cocktail effect \cite{MIRACLE2017448}. HEAs frequently exhibit exceptional functional properties, including superior mechanical strength, corrosion and wear resistance, and remarkable thermal stability, with properties often remaining consistent across a wide temperature range \cite{MIRACLE2017448,SENKOV2011698,STASIAK2022128987}.

The concept of high-entropy materials has also been extended to ceramics such as nitrides, oxides, carbides, and borides \cite{oses2020high}. In these systems, high entropy is typically achieved on the metallic sublattice alone; hence, they are more accurately referred to as high-entropy metallic-sublattice ceramics. These materials incorporate at least five different elements on the metallic sublattice. High-entropy ceramics often adopt a face-centred cubic (fcc) structure. However, it should be noted that certain elemental combinations may favour the formation of amorphous rather than crystalline phases. This behaviour depends on factors such as atomic size mismatch and differences in electronegativity among the constituent elements \cite{VRANA2025131742}.

High-entropy materials can be synthesised both in bulk and thin-film form. For the latter, magnetron sputtering is the most widely used technique \cite{gromov2021structure}, owing to its versatility and ability to achieve far-from-equilibrium conditions. This enables the formation of metastable phases, or even phases unattainable in bulk synthesis \cite{Braun2015}.
High-entropy nitrides based on refractory metals exhibit significant application potential due to their high hardness and thermal stability, attributes inherited from the binary nitrides of their constituent elements. Indeed, hard and stiff refractory-metal-based high-entropy nitrides have already been successfully fabricated \cite{STASIAK2022128987}. However, the actual thermal stability of such thin-film nitrides remains largely unexplored and is currently inferred rather than experimentally confirmed. Existing studies on their temperature stability are predominantly application-driven and confined to relatively moderate temperature regimes—below $600$\textdegree C for solar cell applications \cite{HE2021}, or below $1000$\textdegree C for use as copper diffusion barriers \cite{GRUBER2023130016}.

In this study, we deposited high-entropy Cr-Hf-Mo-Ta-W-N and medium-entropy Mo-Ta-W-N coatings via magnetron sputtering from segmented targets, across a range of chemical compositions and at both ambient temperature and an elevated temperature of 580\textdegree C. The selected compositions were guided by \textit{ab initio} calculations to investigate the role of each element in promoting the formation of high-entropy coatings during deposition. The calculated mixing entropies on the metallic sublattice lie within the range of 1.09--1.61\,R.
Following deposition, the coatings were annealed at 1000\textdegree C and 1200\textdegree C, and their oxidation was examined at 1400\textdegree C. All as-deposited and annealed samples were characterised in terms of chemical composition, morphology, crystalline structure, and mechanical properties. This comprehensive analysis enabled us to assess the influence of individual elements on the thermal stability of the coatings and, with support from \textit{ab initio} modelling, to evaluate the extent of entropy stabilisation at temperatures $\geq$ 1000\textdegree C.

\section{Material and methods}
\subsection{Deposition}
The Cr-Hf-Mo-Ta-W-N coatings were deposited by reactive DC magnetron sputtering from segmented elemental targets in a nitrogen–argon atmosphere. Circular elemental targets (thickness 3\,mm, diameter 50.8\,mm; Cr 99.95\% purity, Mo 99.95\% purity, Ta 99.90\% purity, W 99.95\% purity, Hf 99.9\% purity) were obtained from Testbourne Ltd. Each target was cut into four equal segments and reassembled into composite 4-segment targets mounted on magnetron heads. The deposition system (HVM Flexilab, HVM Plasma s.r.o., Czech Republic) features three magnetron heads in a confocal arrangement.
Two distinct target configurations were employed to produce coatings near equimolar composition and to explore the surrounding compositional space. The first configuration consisted of Cr/Hf (1:3), Ta, and Mo/W (2:2) targets; the second utilised Mo/Hf (2:2), Ta, and Cr/W (2:2). For both setups, the magnetron powers were adjusted to achieve the desired elemental ratios.
Depositions were carried out on two (100)-oriented silicon substrates (2\,$\times$\,1\,cm; ON Semiconductor Czech Republic, s.r.o.) and one R-plane sapphire substrate (1\,$\times$\,1\,cm; Cryscore Optoelectronic Limited, China). Prior to deposition, substrates were ultrasonically cleaned in acetone followed by isopropanol for 180 seconds and mounted at the centre of the rotating substrate holder.
The chamber base pressure of 1\,$\times$\,10$^{-4}$\,Pa was achieved by using an oil-free scroll vacuum pump (Anest Iwata, Japan) and a Hipace 300 turbomolecular pump (Pfeiffer, Germany). Directly before deposition, the substrates were subjected to argon ion bombardment at a pressure of 2.00\,Pa using a 13.56\,MHz-generated DC self-bias of --180\,V for 15 minutes to remove surface contamination and native oxides. The targets, shielded by shutters, were subsequently cleaned with argon ions at 1.26\,Pa for 15 minutes using the same power settings as employed during deposition. Each deposition lasted 45 minutes.
Substrates were rotated at 5 revolutions per minute during deposition. The distance between the targets and the substrate holder was maintained at 110\,mm. Argon flow was fixed at 80 standard cubic centimetres per minute (sccm), while nitrogen flow was set at 20 sccm. The nitrogen flow was chosen to ensure saturation of nitrogen within the coatings, as established in previous work \cite{STASIAK2022128987}. The corresponding total process pressure was 1.5 - 1.6\,Pa.
Samples were deposited at ambient temperature (AT; without intentional heating, $\lesssim$\,50\textdegree C) and at high temperature (HT; 580\textdegree C). For HT samples, the substrate temperature was stabilised for 1\,h prior to deposition to ensure consistent process conditions.

\subsection{Coating characterisation} 
Scanning electron microscopy (SEM) observations of the top-view and cross-section of the deposited coatings were performed using a TESCAN MIRA 3 microscope (TESCAN, Czech Republic). The microscope is equipped with an energy-dispersive X-ray spectroscopy (EDX) detector (X-MAX$^{50}$, Oxford Instruments, UK) for chemical composition analysis. Quantification was carried out using factory standards provided in the AZtec library.
The deposition rate was calculated from thickness measurements obtained from multiple cross-sectional SEM images taken at different positions across the fractured samples.
X-ray diffraction (XRD) measurements were conducted using a Rigaku SmartLab diffractometer with a copper K$\alpha$ radiation source ($\lambda$ = 1.54056\,{\AA}). XRD data analysis was performed using Rigaku PDXL software. Crystallite sizes were estimated using the Scherrer equation \cite{Scherrer1918}.
Mechanical properties of samples on silicon substrates were analysed by Hysitron TI 950 TriboIndenter (Bruker, USA) equipped with a diamond Berkovich indenter tip with a radius of $\sim$50 nm. A certified fused silica sample (Bruker) was used to perform the diamond tip calibration. A matrix of 4 $\times$ 4 standard quasistatic indents, each consisting of 20 partial unloading segments, was carried out in the load-controlled regime with a constant loading rate of 0.2\,mN/s and a maximum load of 11\,mN. The mechanical properties were estimated for the displacement region from 40 nm to 10 \% of film thickness. The indentation data were analysed using the Oliver and Pharr method \cite{oliver1992improved} to determine hardness (H) and the effective Young's modulus (E).
Focused ion beam (FIB) lamellae for transmission electron microscopy (TEM) investigations were prepared using a Thermo Scientific Scios 2 DualBeam microscope. Final thinning steps were performed at 2\,kV. TEM imaging was carried out using a C$_\textrm{s}$-corrected Themis microscope (Thermo Fisher Scientific) operating at 200\,kV. High angle annular dark field (HAADF) images and EDX elemental maps were acquired in scanning transmission electron microscopy (STEM) mode. Selected area electron diffraction (SAED) patterns were analysed using the Process Diffraction software \cite{labar2012electron}. 
To assess the thermal stability of the coatings’ structure and properties, annealing was carried out using a Rapid Thermal Processor (Annealsys AS-ONE 100). The annealing experiments were performed on Si substrates under vacuum, at pressures ranging from 2$\times$10$^{-3}$\,Pa to 2$\times$10$^{-2}$\,Pa. Samples were heated to 1000\textdegree C and 1200\textdegree C with a ramp rate of 10\textdegree C/min, directly followed by cooling at 30\textdegree C/min. Coatings were analysed after cooling to room temperature.
Oxidation was tested using samples deposited on the sapphire substrates. These samples were heated to 1400\textdegree C in the same setup. Upon reaching the maximum temperature, ambient air was partially introduced, raising the pressure from the base level of 2$\times$10$^{-2}$\,Pa to 1$\times$10$^{1}$\,Pa for a duration of 3 minutes. Then, the system was again pumped down, and the samples were left to cool down under vacuum.

\subsection{Ab initio calculations} 

First-principles calculations were carried out based on the projector augmented wave (PAW) method using the Vienna {\it{Ab initio}} Simulation Package (VASP)~\cite{VASP-1,VASP-2} package. 
The exchange-correlation functional was approximated using the Perdew-Burke-Ernzerhof functional revised for solids (PBEsol)~\cite{PBEsol}, with a plane-wave cutoff energy of 600~eV.
The forces on all atoms converged to be less than $10^{-5}$ eV/\AA, and a \textbf{k}-mesh of $7 \times 7 \times 7$ was used for sampling the Brillouin zone, consistently with our previous study~\cite{stasiak2024synthesis}.

The $2 \times 2 \times 2$ fcc supercell (64 atoms, space group Fm$\overline{3}$m) served as a model for high-entropy nitride. Five transition metal (TM) elements, TM$=$(W, Ta, Hf, Mo, Cr), were distributed on the metallic sublattice using the Special Quasirandom Structure (SQS) method~\cite{SQS-1,SQS-2}. Similarly, N vacancies on the non-metallic (N) sublattice were also distributed using SQS at concentrations of 0\%, 25\%, and 50\%.  
For reference, we also modelled binary (TM)N systems, with a fully occupied metallic and N sublattice. 
All supercells were fully relaxed in terms of volume, shape, and atomic positions. 

Chemical stability was assessed based on the formation energy, $E_\mathrm{f}$\cite{koutna2024phase},
\begin{equation}
E_\mathrm{f}=\frac{1}{\sum_s n_s}\left(E_{\mathrm{tot}}-\sum_s n_s \mu_s\right)
\end{equation}
where $E_\mathrm{tot}$ is the total energy of the simulation cell, $n_{s}$ and $\mu_{s}$ are the number of atoms and the chemical potential, respectively, of a species $s$, considering the following structures: bcc-Cr, bcc-Mo, hcp-Hf, bcc-Ta, bcc-W, and N$_{2}$ molecule.
The lattice parameters of the equilibrated supercells were determined as averaged values in the [100], [010], and [001] directions (due to compositional disorder, these were not exactly equal, as expected for an ideal cubic symmetry).

The stress-strain method was used to compute the 4$^\mathrm{th}$-order elasticity tensors of high-entropy and pure binary nitrides \cite{le2001symmetry,koutna2021high}. In Voigt notation, these tensors were transformed into symmetric $6\times6$ matrices, yielding three independent elastic constants ($C_{11}$, $C_{12}$, $C_{44}$) for the cubic system. The polycrystalline Young’s modulus, $E = \frac{9BG}{3B+G}$, was determined using Hill’s average of the bulk and shear moduli (B, G, respectively)~\cite{nye1985physical}, while the Cauchy pressure (used as a ductility indicator) was calculated as $C_{12}-C_{44}$. 

\section{Results and discussion}

\subsection{Composition space mapping}

\subsubsection{Stability trends and structural properties based on ab initio predictions}

\begin{figure}[ht!]
        \centering
        \includegraphics[width=1.0\textwidth]{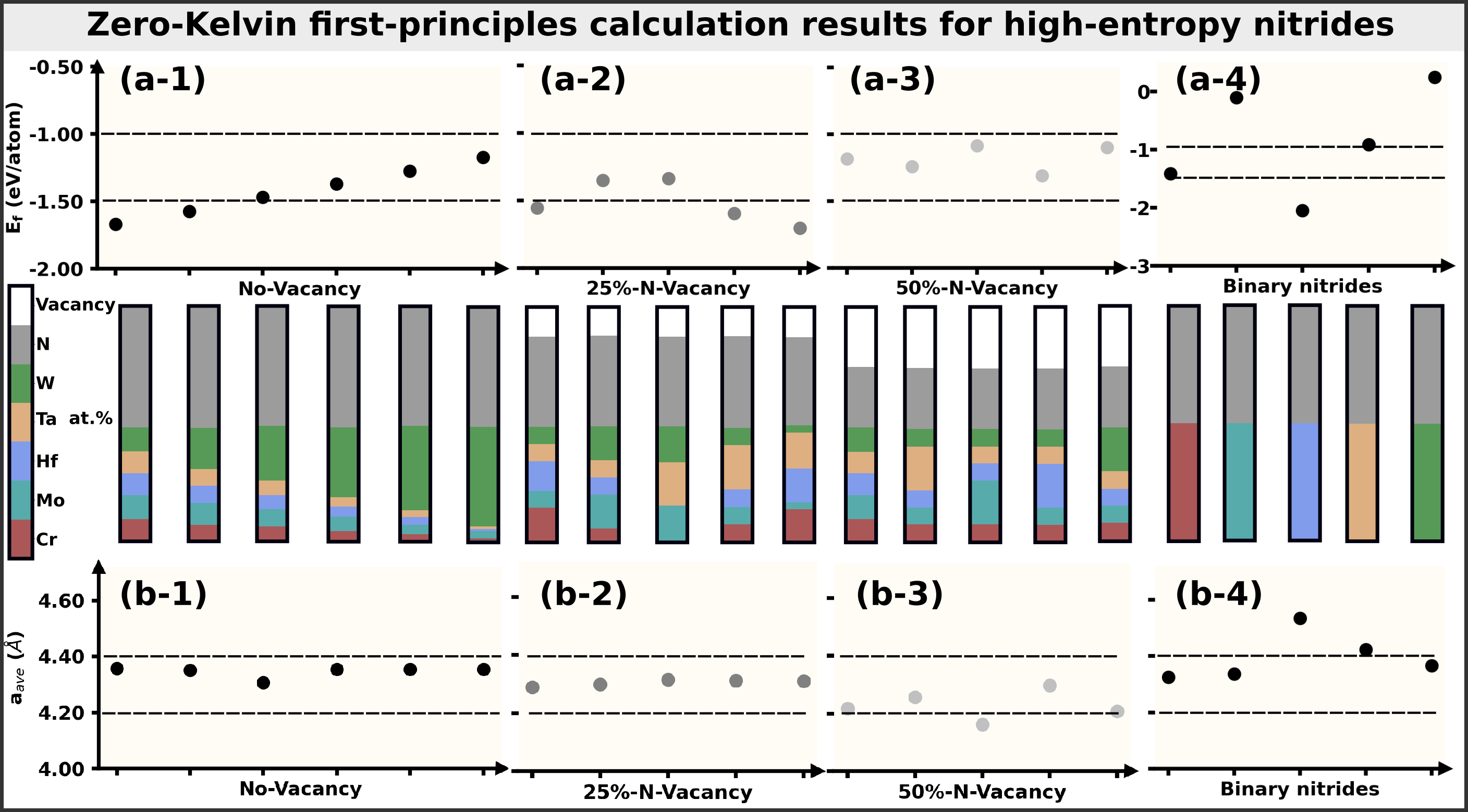}\\
        \caption{DFT-predicted chemical stability (top row, a-*) and structural parameters (bottom row, b-*) of high-entropy nitrides, (W, Ta, Hf, Mo, Cr)N$_x$, with various elemental compositions, together with reference data for the parent binary nitrides. The content of each element is visualised by coloured bars, with gray for N, green for W, yellow for Ta, blue for Hf, cyan for Mo, red for Cr, and white denoting vacancies on the N sublattice. Panels (a) and (b) depict formation energy, $E_\mathrm{f}$, and lattice parameter, $a$ for HENs with fully occupied N sublattice (a-1; b-1); 25\% of N vacancies (a-2; b-2); 50\% of N vacancies (a-3; b-3); and the binary nitrides (a-4; b-4).
        }
        \label{Calc}
\end{figure}

Density functional theory (DFT) calculations were performed to shed light on stability trends in fcc-structured (W, Ta, Hf, Mo, Cr)N$_x$ systems as a function of their elemental composition, including off-stoichiometry on the N sublattice (Figure~\ref{Calc}). 
The energy of formation, $E_\mathrm{f}$, served as a descriptor of chemical stability, with more negative values indicating more stable structures.
Additionally, note that all HENs presented in Figure~\ref{Calc} satisfy mechanical stability conditions, based on their elastic constants relations~\cite{mouhat2014necessary}. 
Among the reference binary nitrides, however, only HfN, TaN, and CrN are mechanically stable in the 1:1 TM:N stoichiometry, while the fcc-structured MoN ($E_\mathrm{f}\approx{-0.10}$~eV/at.) and WN ($E_\mathrm{f}\approx{0.25}$~eV/at.) are energetically and mechanically stabilised by N and/or TM vacancies~\cite{koutna2016point,balasubramanian2016vacancy,lee2021defect}, with most commonly reported compositions being MoN$_{0.5}$~\cite{klimashin2016impact,ozsdolay2017cation} and WN$_{0.5}$~\cite{ozsdolay2016cubic,buchinger2019toughness}.

Figure~\ref{Calc}-(a/b-1) illustrates how changes in the content of W---presumably the most destabilising element in view of its metastability/instability at 1:1 stoichiometry---affect formation energy (a-1) and lattice parameter (b-1) of HENs with a fully occupied N sublattice.
As the W fraction increases from ideally equimolar up to $\approx{40}$\%, the corresponding $E_\mathrm{f}$ increases significantly, by $\approx{30}$\%, pointing towards limited stability of W-rich coatings. Throughout increasing the W content, the lattice parameter remains nearly unaffected, fluctuating by less than 1.15\%.

Next, the role of nitrogen vacancies on the stabilisation of the fcc phase is investigated. HENs with 25\% N substoichiometry, Figure~\ref{Calc}-(a-2), exhibit fairly low $E_\mathrm{f}$ values.
The most stable variants are those with higher Hf, Ta, and Cr contents (14.06\% each, and 6.26\% for Mo and W together), consistently with these elements, particularly Hf, being strong nitride formers and preferably crystallising in near-stoichiometric or medium N sub-stoichiometric compositions~\cite{koutna2016point,zhang2013insights,koutna2021high}.  
Furthermore, Ta-, Hf-, and Cr-rich HENs yield $E_\mathrm{f}$ comparable with or even below that of the parent perfect equimolar HENs, suggesting that up to 25\% vacancies on the N sublattice can have a stabilising effect. 
Lattice parameters of HENs with 25\% of N vacancies, Figure~\ref{Calc}-(b-2), are $\approx{2}\%$ lower than those of vacancy-free HENs. 

A further increase of N vacancy concentration, up to 50\% of the non-metallic sublattice is followed by an expected $E_\mathrm{f}$ increase (Figure~\ref{Calc}-(a-3)) in combination with the lattice parameter shrinkage (Figure~\ref{Calc}-(b-3)). 
Again, as Hf is the strongest nitride former, increasing its contents (up to $\approx{19}$~\%, with other elements reaching $\approx{8}$~\% each) has a stabilising effect.
Similar stabilising effects may be achieved by Ta.
While Ta certainly does not belong to as strong nitride formers as the group 4 transition metals, its tolerance for wide stoichiometry changes---including fairly high N sub-stoichiommetry~\cite{koutna2016point}---may be beneficial, especially as both Mo and W exhibit a high driving force for vacancies.
The Mo-rich structure shows the lowest lattice parameter (4.16~{\AA}, close to 4.21~{\AA} of MoN$_{0.5}$~\cite{koutna2016point}), while other configurations yield lattice parameters $\approx{4.25}$~{\AA} which is $\approx{2}$\% decrease with respect to HENs with 25\% N vacancy contents.

In summary, DFT calculations indicated that HENs combining Hf, Ta, Cr, Mo, and W can be stable in the cubic fcc structure, with $E_\mathrm{f}=-1.68$~eV/at. and $a=4.40$~{\AA} for the equimolar compositions with fully occupied N sublattice. 
This formation energy is $\approx{0.2}$~eV/at. above that of the strongest nitride formers, TiN, ZrN, HfN, and $\approx{0.6}$~eV/at. below the ``rule of mixture'' for the parent nitrides, pointing towards significant stabilisation effect of entropy and lattice distortions.
Consistently with stability trends for the parent binary nitrides, higher chemical stability can be achieved in Hf-rich (Ta-, Cr-) HENs. 
Furthermore, the fcc-(Hf, Ta, Cr, Mo, W)N$_x$ is predicted to form especially upon the introduction of N vacancies, where especially contents around 25~\% may be favoured, as fcc-MoN and fcc-WN are chemically and/or mechanically stabilised as N-vacant, and fcc-TaN tolerates wide stoichiometry changes.

\subsubsection{Chemical composition}

The initial target powers were set to achieve a nearly equimolar composition of the metallic elements. As the substrate temperature was found to have no significant influence on the resulting chemical composition, identical power settings were used for both the ambient temperature (AT) and high temperature (HT) series. Subsequently, the target powers were systematically adjusted to investigate the influence of individual elements on the properties of the coatings.

The specific target configurations, applied powers, resulting chemical compositions, and calculated mixing entropies on the metallic sublattice are summarised in Table~\ref{tab:depo}. Samples AT1 and HT1, deposited at ambient and elevated temperature respectively, exhibit a nearly equimolar metallic composition. Samples AT2 and HT2 are deficient in Cr and Hf; AT3 and HT3 are Ta-deficient; AT4 and HT4 are deficient in Mo and W. Conversely, samples AT5 and HT5 are enriched in Cr and Hf; AT6 and HT6 are Ta-rich; and AT7 and HT7 are rich in Mo and W. Samples AT8 and HT8 were prepared without any Cr or Hf. Samples AT9 and HT9 are enriched in W and Cr, while samples AT10 and HT10 are Mo-rich.

The Cr-Hf-rich and Cr-Hf-free compositions were selected based on \textit{ab initio} calculations, which identified the Cr-Hf-enriched phase as having the lowest formation energy of --1.704\,eV/atom. The highest calculated mixing entropy on the metallic sublattice, approximately 1.61\,R, corresponds to the nearly equimolar coatings AT1 and HT1. The lowest mixing entropy, around 1.1\,R, was obtained for coatings AT8 and HT8, which lack both chromium and hafnium. The second-lowest entropy, $\approx{1.3}$\,R, is associated with samples AT9 and HT9, which are rich in chromium and tungsten. The remaining coatings exhibit mixing entropies in the range of 1.5--1.6\,R.

\begin{table}[htpb]
\small
\centering
\hspace*{-1.2cm}
\begin{tabular}{c | c c c c c | c c c c c c | c}       
sample & \multicolumn{5}{c}{power [W]} & \multicolumn{6}{c}{chemical composition [at.\%]} & Mixing\\
 & Cr/Hf & Ta & Mo/W & Mo/Hf & Cr/W & N & Cr & Mo & Hf & Ta & W & entropy [R]\\
 & (1:3) &  & (2:2) & (2:2) & (2:2) &  &  &  &  &  &  &\\
 \hline
AT1 & 150 & 75 & 70 &  &  & 48.1 & 10.1 & 10.1 & 9.6 & 10.7 & 11.3 & 1.61\\
AT2 & 75 & 75 & 70 &  &  & 48.6 & 5.9 & 13.4 & 4.7 & 14.0 & 13.4 & 1.52\\
AT3 & 150 & 37 & 70 &  &  & 48.1 & 11.6 & 11.8 & 11.0 & 5.1 & 12.4 & 1.57\\
AT4 & 150 & 75 & 35 &  &  & 51.4 & 12.2 & 5.8 & 11.6 & 13.2 & 5.8 & 1.55\\
AT5 & 150 & 37 & 35 &  &  & 49.1 & 15.1 & 6.9 & 14.1 & 6.6 & 8.2 & 1.55\\
AT6 & 150 & 150 & 70 &  &  & 50.2 & 7.2 & 7.6 & 7.0 & 19.8 & 8.2 & 1.51\\
AT7 & 150 & 75 & 140 &  &  & 44.0 & 7.7 & 16.7 & 7.2 & 8.1 & 16.4 & 1.54\\
AT8 & 0 & 150 & 140 &  &  & 43.9 & 0.0 & 17.6 & 0.0 & 20.8 & 17.6 & 1.1\\
AT9 &  & 75 &  & 60 & 150 & 44.6 & 21.9 & 5.4 & 1.3 & 7.7 & 19.2 & 1.32\\
AT10 &  & 75 &  & 150 & 60 & 47.0 & 8.6 & 20.1 & 5.5 & 9.7 & 9.0 & 1.51\\
 &  &  &  &  &  &  &  &  &  &  &  & \\
HT1 & 150 & 75 & 70 &  &  & 45.6 & 9.9 & 10.6 & 9.5 & 11.5 & 12.9 & 1.60\\
HT2 & 75 & 75 & 70 &  &  & 41.8 & 6.1 & 14.9 & 4.7 & 15.8 & 16.7 & 1.50\\
HT3 & 150 & 37 & 70 &  &  & 44.2 & 12.0 & 12.7 & 11.5 & 5.7 & 13.9 & 1.57\\
HT4 & 150 & 75 & 35 &  &  & 47.3 & 11.6 & 6.6 & 12.1 & 15.1 & 7.5 & 1.56\\
HT5 & 150 & 37 & 35 &  &  & 47.2 & 14.8 & 7.8 & 14.4 & 7.2 & 8.7 & 1.56\\
HT6 & 150 & 150 & 70 &  &  & 46.9 & 7.0 & 8.4 & 7.1 & 22.5 & 8.1 & 1.48\\
HT7 & 150 & 75 & 140 &  &  & 40.6 & 7.6 & 17.8 & 7.3 & 9.0 & 17.7 & 1.54\\
HT8 & 0 & 150 & 140 &  &  & 37.0 & 0.0 & 19.6 & 0.0 & 24.6 & 18.7 & 1.09\\
HT9 &  & 75 &  & 60 & 150 & 37.8 & 24.6 & 6.5 & 1.3 & 8.9 & 21.0 & 1.33\\
HT10 &  & 75 &  & 150 & 60 & 43.4 & 9.5 & 21.5 & 5.8 & 10.4 & 9.4 & 1.51\\
\end{tabular}
\caption{Deposition parameters, chemical composition measured by EDX, and the mixing entropy on the metallic sublattice calculated from the chemical composition.}
\label{tab:depo}
\end{table}

The nitrogen content of samples AT1--AT6 and AT10 is relatively consistent, at approximately 49\,at.\%. In contrast, samples AT7--AT9 exhibit significantly lower nitrogen content, accompanied by a markedly higher tungsten content compared to the other AT coatings. This behaviour arises from the tungsten target becoming poisoned at higher nitrogen flow rates than other elemental targets. Consequently, increasing the power supplied to the tungsten target keeps it within the metallic sputtering regime, leading to higher sputter yields not fully compensated by the available nitrogen flow \cite{SAFI2000203}.
Similar trends were observed in the HT series. However, the overall nitrogen content in these samples is lower, a phenomenon frequently reported in the literature. This reduction is attributed to the enhanced desorption of nitrogen atoms from non-equilibrium lattice sites due to the increased energy at elevated substrate temperatures \cite{STASIAK2022128987,LIANG20117709,VONFIEANDT2020137685}.

The measured chemical compositions suggest the presence of vacancies on the nitrogen sublattice in the majority of the samples. This observation aligns with both the \textit{ab initio} calculations and previous findings indicating that most binary nitrides within the studied system are stabilised by nitrogen vacancies \cite{koutna2016point,balasubramanian2016vacancy,lee2021defect}.

\subsubsection{Morphology and microstructure}

\begin{figure}[htpb]
        \centering
        \includegraphics[width=.9\textwidth]{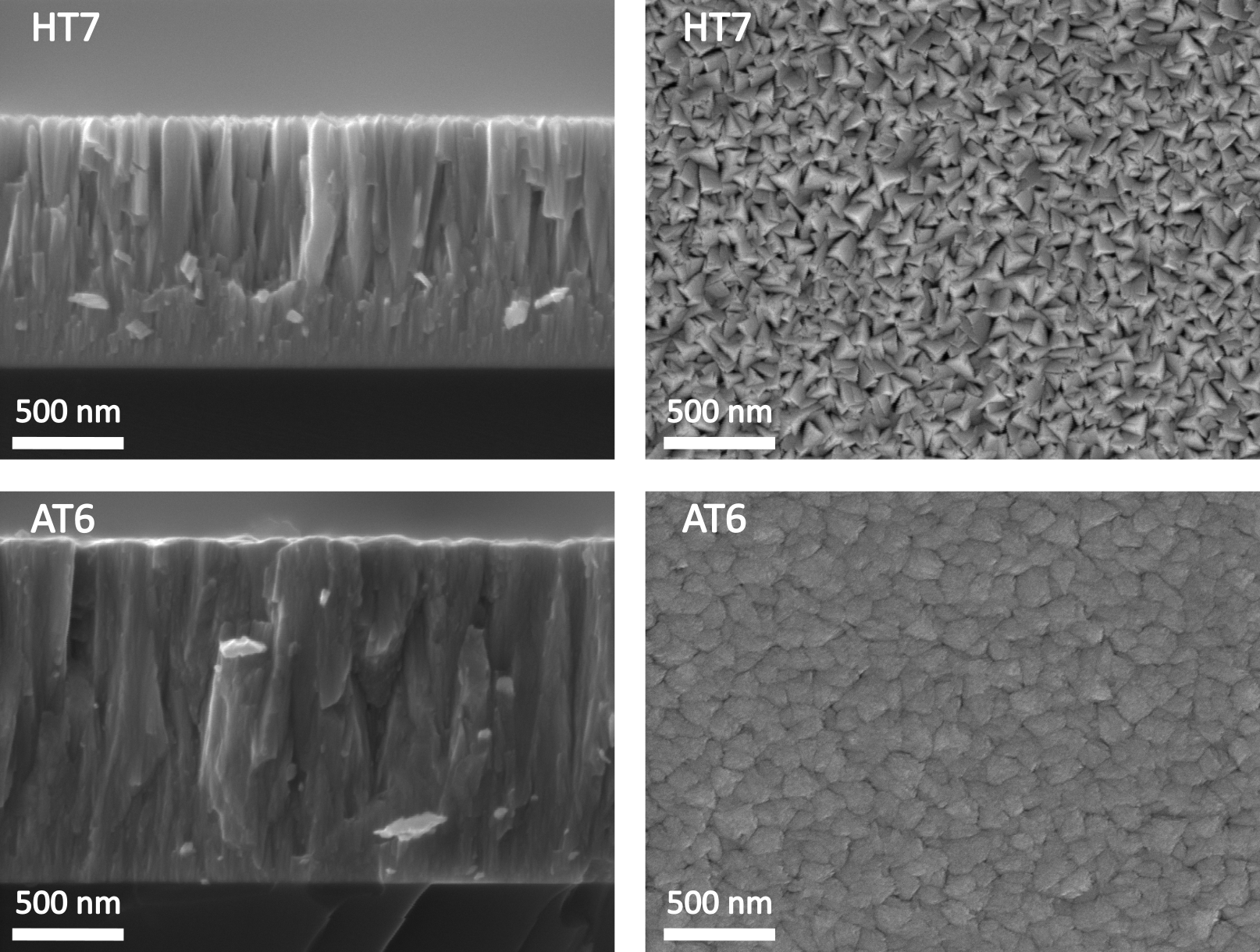}\\
        \caption{SEM micrographs showing typical morphology of the deposited coatings.}
        \label{semy1}
\end{figure}

SEM micrographs showing the typical morphology of the deposited coatings are presented in Figure~\ref{semy1}. 
SEM micrographs of unpolished fracture cross-sections of all AT and HT samples are provided in Figures~S1 and S2, respectively, in the Supplementary Material.
All coatings exhibit a V-shaped columnar growth, which is characteristic of zone T of the materials' structure zone models \cite{ANDERS20104087}. The column widths range from approximately 200 to 500\,nm. Most coatings display well-defined triangular surface features, as illustrated for sample HT7 in Figure~\ref{semy1}, with feature sizes corresponding closely to the underlying column widths. Such surface morphology is often indicative of a strong (111) texture in the growth direction of a cubic crystalline phase.
Samples AT6, AT8, and HT4, however, exhibit a more cauliflower-like morphology, which is also commonly observed in magnetron-sputtered films \cite{ANDERS20104087}. This variation in surface morphology is likely linked to a combination of coating chemistry and deposition energetics. Specifically, AT6, AT8, and HT4 have higher tantalum content, and the deposition powers influence the flux of the arriving species.

\begin{figure}[htpb]
        \centering
        \includegraphics[width=1\textwidth]{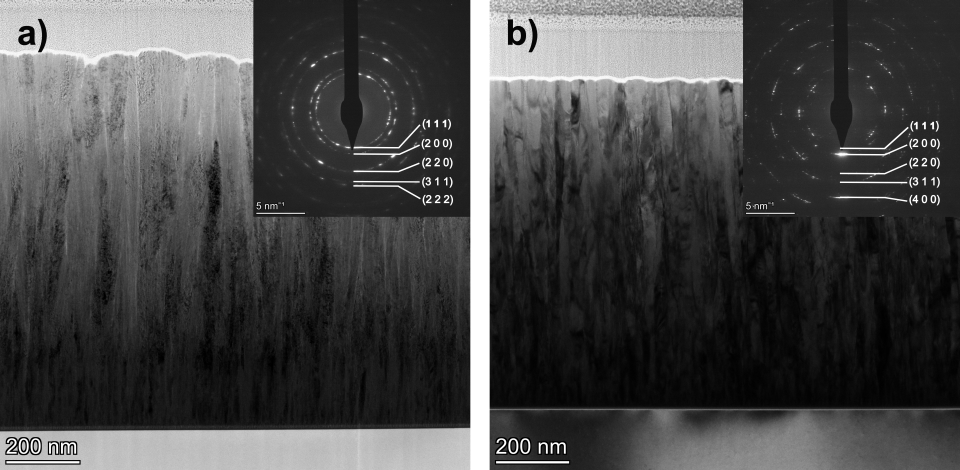}\\
        \caption{TEM micrographs of coatings a) AT4 and b) HT4 showing typical microstructure of the deposited coatings. SAED patterns with rings identified as fcc structure are plotted as an inset in each micrograph.}
        \label{temy1}
\end{figure}

\begin{figure}[htpb]
        \centering
        \includegraphics[width=1\textwidth]{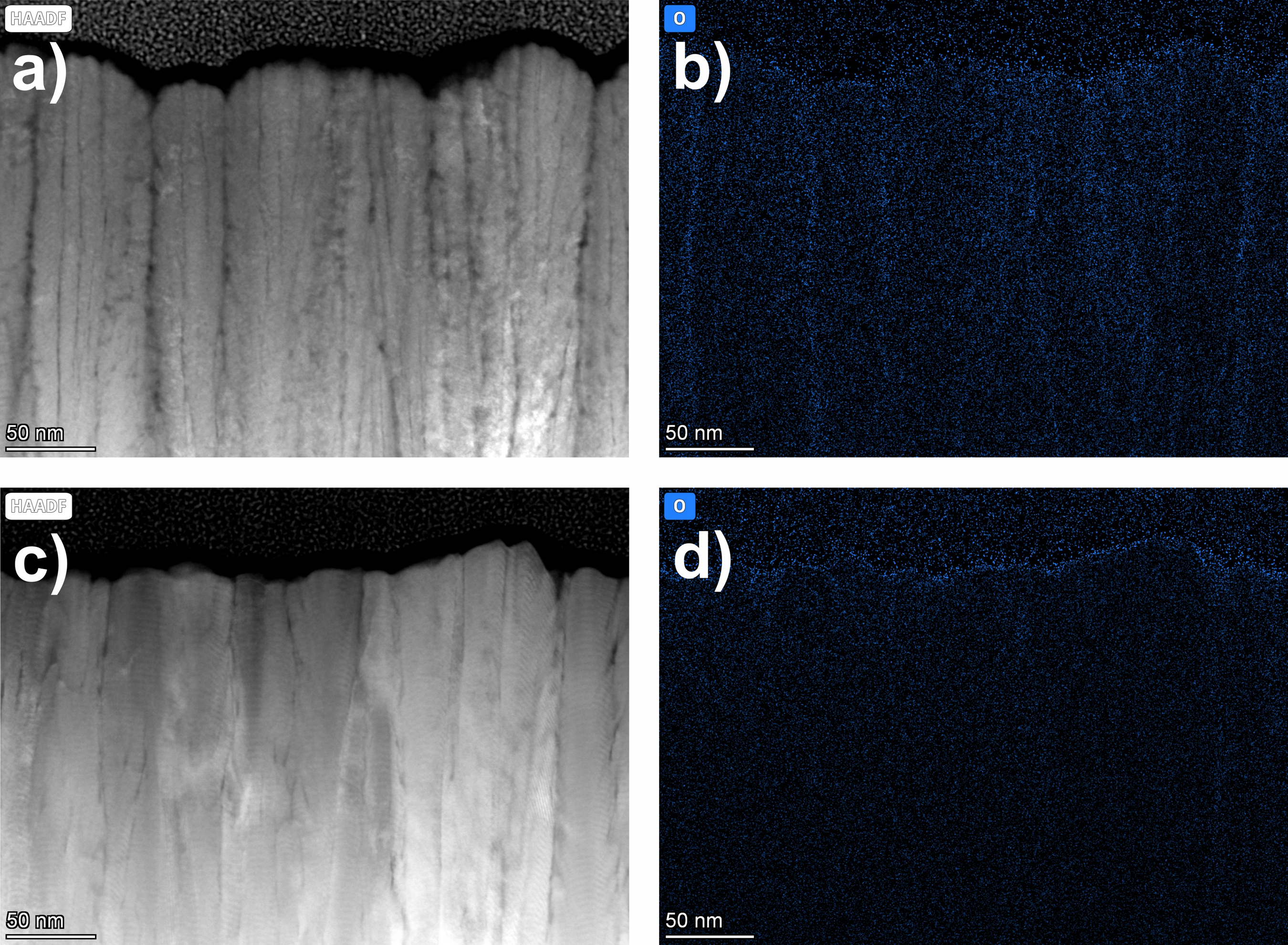}\\
        \caption{STEM HAADF micrographs of the upper part of samples a) AT4 and c) HT4. corresponding EDX oxygen maps are shown in b) and d).}
        \label{temy2}
\end{figure}

The columnar structure is confirmed by TEM images of samples AT4 and HT4, shown in Figure~\ref{temy1}. The column widths observed via TEM are narrower—ranging from 50 to 200\,nm in both films—indicating that the broader columns visible in SEM images are composed of multiple finer sub-columns. Sample HT4 exhibits a dense microstructure, whereas AT4 shows columns separated by 1--2\,nm wide amorphous boundaries. The oxygen elemental map in Figure~\ref{temy2} reveals that these boundaries are enriched in oxygen. Such oxygen-rich column boundaries may act as structural defects, potentially reducing hardness and serving as initiation sites for oxidation at elevated temperatures. Selected area electron diffraction (SAED) patterns of the AT4 and HT4 samples, shown as insets in Figure~\ref{temy1}, confirm the presence of a face-centred cubic (fcc) structure with a lattice parameter of 4.34\,\AA.
 
The rotation of the substrate holder induces a layered structure that is most distinct in the initial $\sim$200\,nm of sample AT4 and $\sim$100\,nm of HT4. As the V-shaped columnar growth develops and the surface layer of the growing coating becomes more rugged, this layered morphology becomes less pronounced, particularly in the HT coatings, where the increased adatom energy and thus enhanced diffusion to lower energy sites promote more even mixing of the elements. The layer contrast is more prominent in the AT series and is illustrated in Figure~S3 in the Supplementary Material, which compares HAADF images from the top and bottom regions of AT4 and HT4 coatings.

\begin{figure}[htpb]
        \centering
        \includegraphics[width=.49\textwidth]{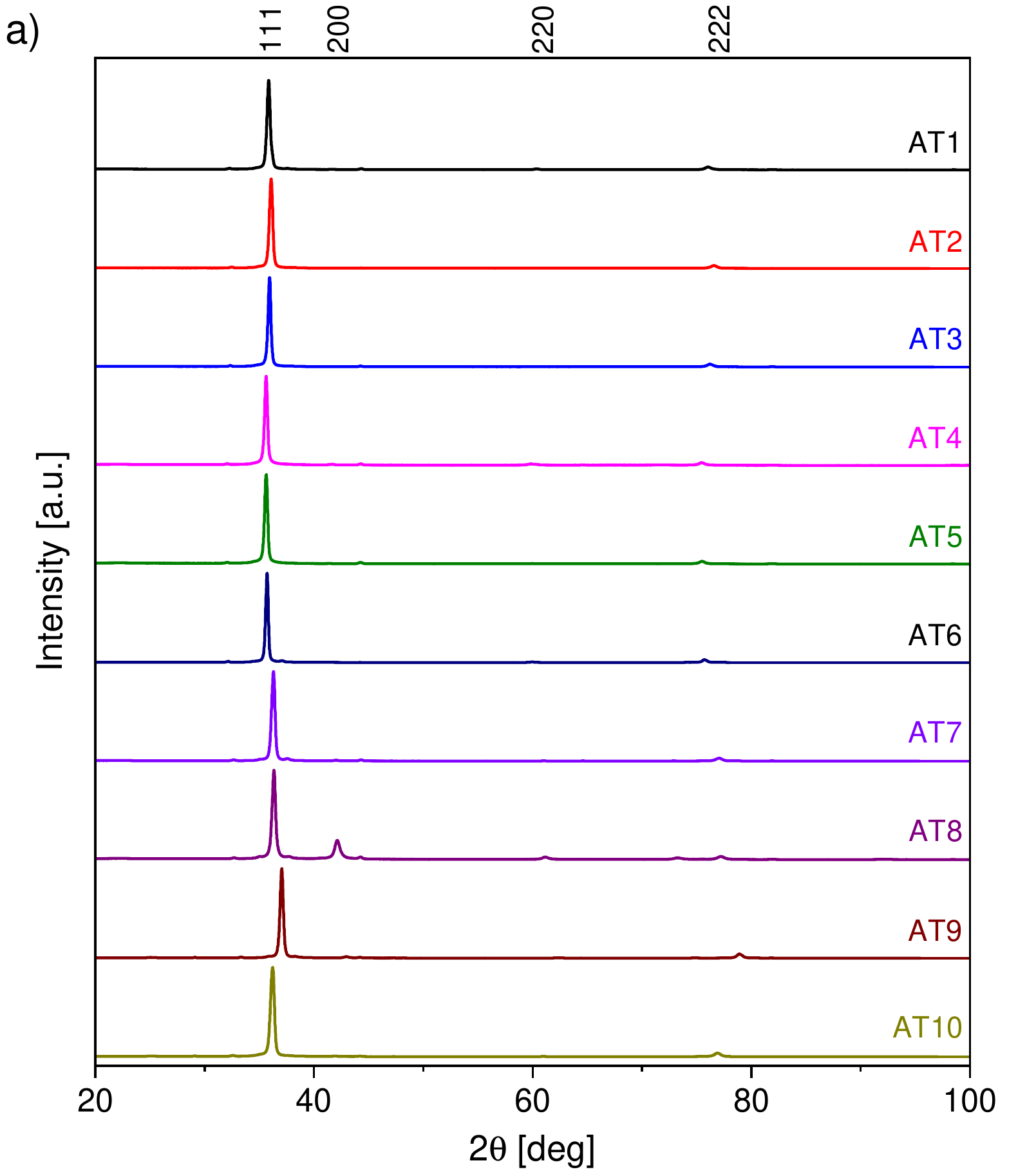}
        \includegraphics[width=.49\textwidth]{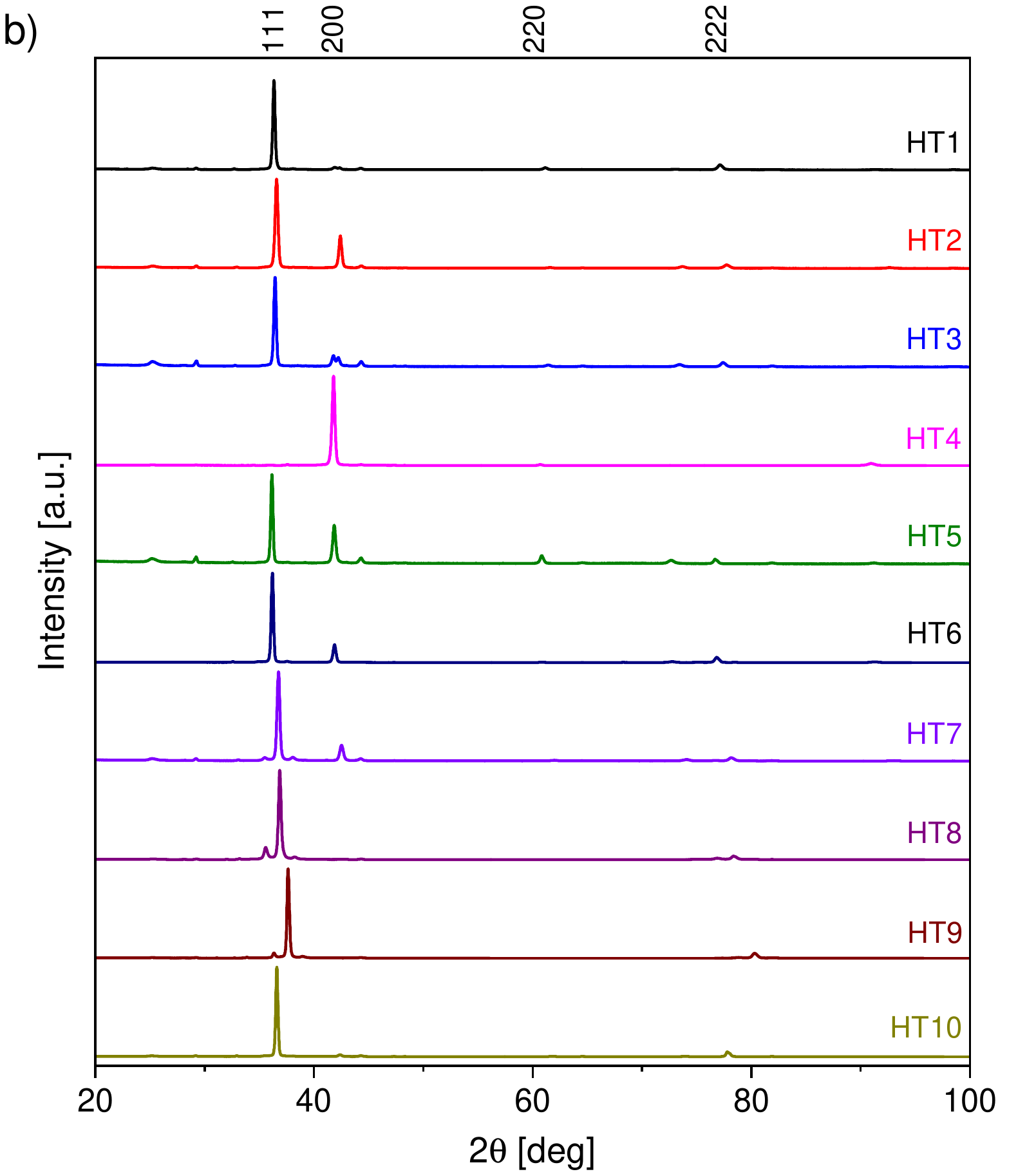}\\
        \caption{X-ray diffractograms of the a) AT and b) HT coatings. Miller indices corresponding to an fcc phase are provided at the top of the image.}
        \label{xrd1}
\end{figure}

X-ray diffractograms of the deposited coatings are presented in Figure~\ref{xrd1}. The positions of the major peaks corresponding to a face-centred cubic (fcc) lattice are indicated by their respective Miller indices above the plots. All coatings exhibit diffraction peaks consistent with an fcc phase. Minor peaks observed in the spectra can be attributed to the substrate, substrate holder, or K$_{\beta}$ radiation artefacts.
These results confirm that all deposited coatings contain an fcc nitride phase. For samples incorporating all five metallic elements, this phase can be classified as a high-entropy nitride phase.

The vast majority of the coatings exhibit a pronounced (111) texture in the growth direction, consistent with the observed surface morphology. Such a (111) texture is frequently reported in face-centred cubic (fcc) refractory-metal-based high-entropy nitrides \cite{STASIAK2022128987,LAI20063275,REN2013171}. This preference arises from the (111) plane’s superior ability to accommodate strain energy \cite{PELLEG1991117}, which is particularly relevant in high-entropy nitride (HEN) coatings due to substantial lattice strain caused by atomic size mismatch.
In our case, depending on the chemical composition, the atomic size difference on the metallic sublattice ranges from approximately 5.5--9\,\% for samples containing all five elements, and about 2.3\,\% for samples AT8 and HT8. When considering the entire lattice, including nitrogen, the atomic size mismatch increases to approximately 26--31.5\,\%. All calculated atomic size differences are summarised in Table~\ref{tab:prop}.
Interestingly, sample HT4—despite showing the second-highest atomic size difference on the metallic sublattice (8\,\%) and the highest overall mismatch in the HT series (30.8\,\%)—exhibits a strong (200) texture rather than (111). This highlights the complexity of the deposition process, where not only thermodynamics but also growth kinetics and deposition energetics play a critical role. While the (111) plane accommodates lattice strain more effectively, the (100) planes exhibit the lowest surface energy \cite{PELLEG1991117}, making their preferential growth possible under certain deposition conditions.

\begin{table}
\small
\centering
\hspace*{-1cm}
\begin{tabular}{c | c c | c c | c c c c}     
sample & ASD$_{\textrm{metal}}$ & ASD$_{\textrm{tot}}$ & a [\AA] & D & H [GPa] & E [GPa] & H/E & H$^3$/E$^2$ \\
 & [\%] & [\%] & & [nm] &  & &  & [GPa] \\
 \hline
AT1 & 7.50 & 30.50 & 4.335 $\pm$ 0.001 &36& 12.8 $\pm$ 1.3 & 219 $\pm$ 11 & 0.058 & 0.044 \\
AT2 & 5.70 & 30.07 & 4.311 $\pm$ 0.002 &41& 11.7 $\pm$ 0.7 & 211 $\pm$ 9 & 0.056 & 0.036 \\
AT3 & 7.99 & 30.47 & 4.328 $\pm$ 0.002 &45& 10.2 $\pm$ 0.9 & 194 $\pm$ 10 & 0.052 & 0.028 \\
AT4 & 8.35 & 31.56 & 4.363 $\pm$ 0.001 &40& 14.0 $\pm$ 1.1 & 220 $\pm$ 10 & 0.063 & 0.056 \\
AT5 & 9.04 & 31.14 & 4.363 $\pm$ 0.001 &41& 11.4 $\pm$ 1.6 & 206 $\pm$ 14 & 0.055 & 0.035 \\
AT6 & 6.55 & 30.99 & 4.355 $\pm$ 0.005 &55& 16.7 $\pm$ 1.4 & 210 $\pm$ 17 & 0.079 & 0.106 \\
AT7 & 6.38 & 29.13 & 4.290 $\pm$ 0.003 &38& 13.3 $\pm$ 1.3 & 231 $\pm$ 13 & 0.058 & 0.044 \\
AT8 & 2.27 & 28.71 & 4.281 $\pm$ 0.002 &31& 15.5 $\pm$ 1.6 & 245 $\pm$ 15 & 0.063 & 0.062 \\
AT9 & 5.78 & 27.53 & 4.203 $\pm$ 0.004 &39& 8.7 $\pm$ 1.6 & 197 $\pm$ 16 & 0.044 & 0.017 \\
AT10 & 6.26 & 29.59 & 4.296 $\pm$ 0.001 &33& 8.7 $\pm$ 1.3 & 192 $\pm$ 13 & 0.045 & 0.018 \\
 &  &  &  &  &  &  &  \\
HT1 & 7.29 & 29.96 & 4.283 $\pm$ 0.003 &71& 14.1 $\pm$ 2.1 & 202 $\pm$ 18 & 0.070 & 0.069 \\
HT2 & 5.44 & 28.52 & 4.256 $\pm$ 0.003 &51& 12.7 $\pm$ 2.4 & 252 $\pm$ 26 & 0.051 & 0.033 \\
HT3 & 7.87 & 29.66 & 4.270 $\pm$ 0.001 &66& 10.7 $\pm$ 1.9 & 211 $\pm$ 20 & 0.051 & 0.027 \\
HT4 & 8.03 & 30.78 & 4.319 $\pm$ 0.003 &110& 19.5 $\pm$ 3.0 & 284 $\pm$ 27 & 0.069 & 0.092 \\
HT5 & 8.88 & 30.76 & 4.306 $\pm$ 0.006 &84& 16.1 $\pm$ 2.5 & 263 $\pm$ 23 & 0.061 & 0.060 \\
HT6 & 6.35 & 30.38 & 4.302 $\pm$ 0.007 &91& 18.1 $\pm$ 3.8 & 263 $\pm$ 37 & 0.069 & 0.085 \\
HT7 & 6.22 & 28.32 & 4.234 $\pm$ 0.001 &56& 10.5 $\pm$ 2.0 & 241 $\pm$ 21 & 0.043 & 0.020 \\
HT8 & 2.31 & 26.92 & 4.219 $\pm$ 0.005 &51& 11.3 $\pm$ 1.5 & 238 $\pm$ 15 & 0.047 & 0.025 \\
HT9 & 5.75 & 25.97 & 4.138 $\pm$ 0.002 &76& 9.9 $\pm$ 1.3 & 232 $\pm$ 13 & 0.043 & 0.018 \\
HT10 & 6.26 & 28.78 & 4.247 $\pm$ 0.002 &91& 7.9 $\pm$ 1.3 & 210 $\pm$ 15 & 0.037 & 0.011 \\
\end{tabular}
\caption{Properties of the deposited coatings: atomic size difference considering the metallic sublattice (ASD$_{\textrm{metal}}$) and considering the whole lattice (ASD$_{\textrm{tot}}$), lattice parameter (a), crystallite size (D), hardness (H), effective elastic modulus (E), H/E and H$^3$/E$^2$ ratios.}
\label{tab:prop}
\end{table}

\begin{figure}[htpb]
        \centering
        \includegraphics[width=.49\textwidth]{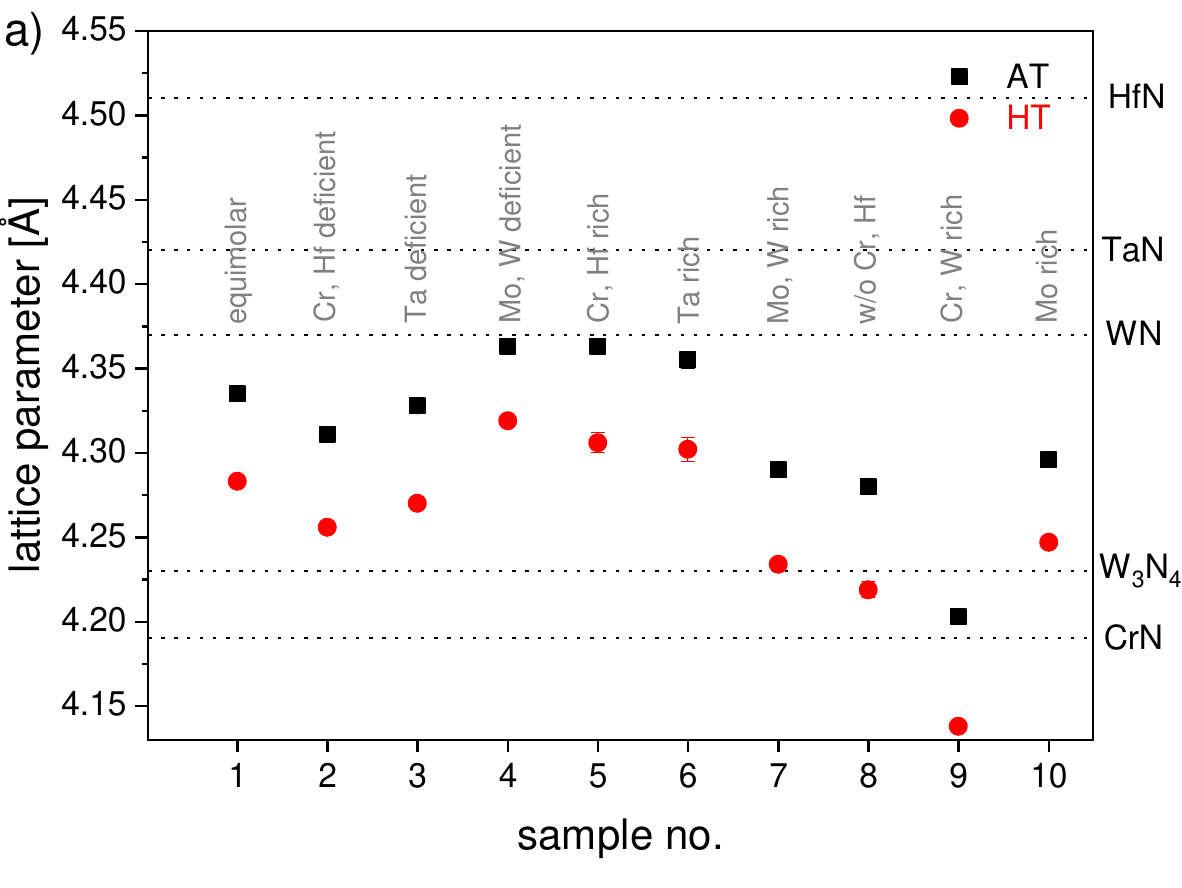}
        \includegraphics[width=.49\textwidth]{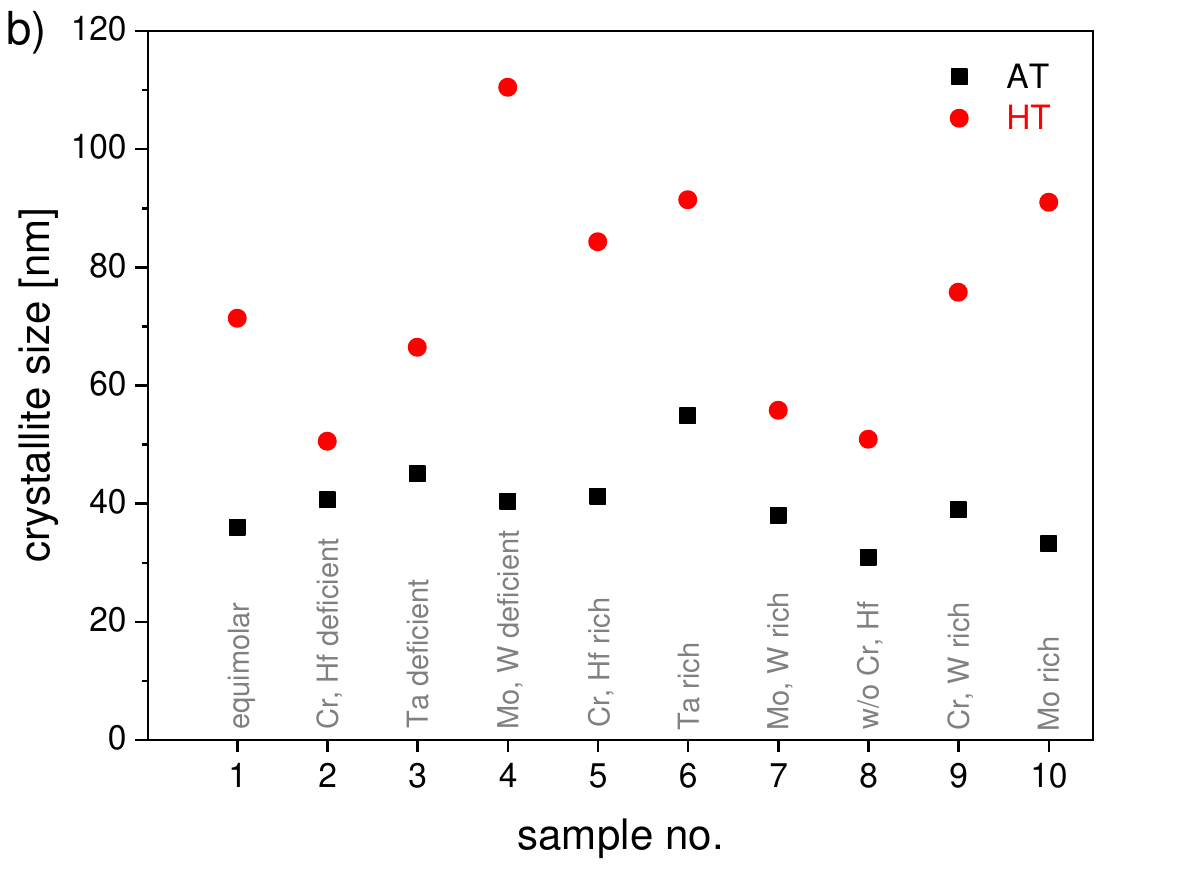}\\
        \caption{a) Lattice parameters of the deposited coatings determined from XRD measurements. Reference values for selected binary nitrides are shown as horizontal dotted lines \cite{matproj}. b) Crystallite sizes determined from the most intensive diffraction peak.}
        \label{latpar1}
\end{figure}

The lattice parameters of the deposited coatings are shown in Figure~\ref{latpar1}a. Reference values for selected binary nitrides are included as horizontal dotted lines. These reference values were taken from the Materials Project database \cite{matproj}, considering cubic and orthorhombic binary nitrides that have been either experimentally observed or predicted to be stable, and whose lattice parameters fall within a range comparable to those of the deposited coatings.
The measured lattice parameters from the most intensive peak are consistently higher for the AT coatings compared to their HT counterparts. This difference can be attributed partly to a lower concentration of nitrogen vacancies in the AT coatings, as indicated by the chemical composition data, and partly to expected higher compressive stresses. The elevated substrate temperature during HT deposition increases adatom mobility, allowing atoms to diffuse to energetically favourable sites more efficiently \cite{KUNC2003744,HERAULT2021117385}, thereby reducing internal stress.
This notion is further supported by comparison with the theoretical lattice parameters plotted in Figure~\ref{Calc}(b). The lattice parameters of the HT coatings are similar to, or slightly larger than, the theoretical values—even accounting for nitrogen vacancies—suggesting the presence of low-to-moderate residual stress. In contrast, the significantly larger lattice parameters of the AT coatings imply higher compressive stress.
Stress at the lattice level was also assessed using the $\sin^2{\psi}$ method. However, the data are inconclusive. The majority of the samples exhibit a V-shaped trend in the $2\theta$--$\sin^2{\psi}$ plots: measurements at low $\sin^2{\psi}$ values (probing deeper regions) indicate tensile stress, whereas those at higher $\sin^2{\psi}$ (near the surface) indicate compressive stress. Notably, both tensile and compressive components are of greater magnitude in HT samples than in AT ones.
It is important to note that lattice-level microstress evaluation via XRD is an indirect method, and the resulting $2\theta$--$\sin^2{\psi}$ plots can be affected by other factors, including elastic anisotropy, preferred orientation (texture), or the presence of coarse grains \cite{birkholz2006thin,sarmast2023residual}.
 
Within both the AT and HT series, the lattice parameter is influenced by the chemical composition of each coating, particularly by the atomic sizes of the metallic elements and the lattice parameters of their corresponding binary nitrides. An increase in tantalum and hafnium content results in a higher lattice parameter, as both their nitrides possess larger lattice constants than the equimolar reference, and their atomic radii—1.43\,\AA\ for Ta and 1.58\,\AA\ for Hf—are above the average metallic atomic radius of 1.40\,\AA.
Conversely, increasing the content of tungsten, chromium, and molybdenum tends to reduce the lattice parameter. Chromium, with the smallest atomic radius of 1.25\,\AA, also forms nitrides with the lowest lattice parameters in the studied set. Tungsten forms two different binary nitrides in the relevant range—WN and W$_3$N$_4$—one with a higher and one with a lower lattice parameter than the equimolar coating. Nevertheless, tungsten’s atomic radius of 1.37\,\AA\ is below the average. Molybdenum, while lacking a stable cubic or orthorhombic nitride with a lattice parameter close to the deposited coatings, has an atomic radius of 1.36\,\AA, which is similarly below the average value.

The crystallite sizes of the coatings are presented in Figure~\ref{latpar1}b. Crystallite size was estimated from the full width at half maximum (FWHM) of the most intense diffraction peak—(200) for sample HT4, and (111) for all other samples. 
In the AT series, crystallite size remained relatively constant, ranging from approximately 30 to 45\,nm. An exception was observed for the Ta-rich sample, which exhibited a larger crystallite size of 55\,nm. All coatings deposited at high temperature (HT series) showed increased crystallite sizes compared to their ambient-temperature counterparts, consistent with the enhanced adatom mobility and resulting crystallite coarsening at elevated temperatures \cite{Levi_1997}. 
Within the HT series, crystallite sizes ranged from approximately 50 to 110\,nm. A trend was observed in which higher contents of hafnium, tantalum, and chromium promoted the formation of larger crystallites.


\subsubsection{Mechanical properties}

\begin{figure}[htpb]
        \centering
        \includegraphics[width=.49\textwidth]{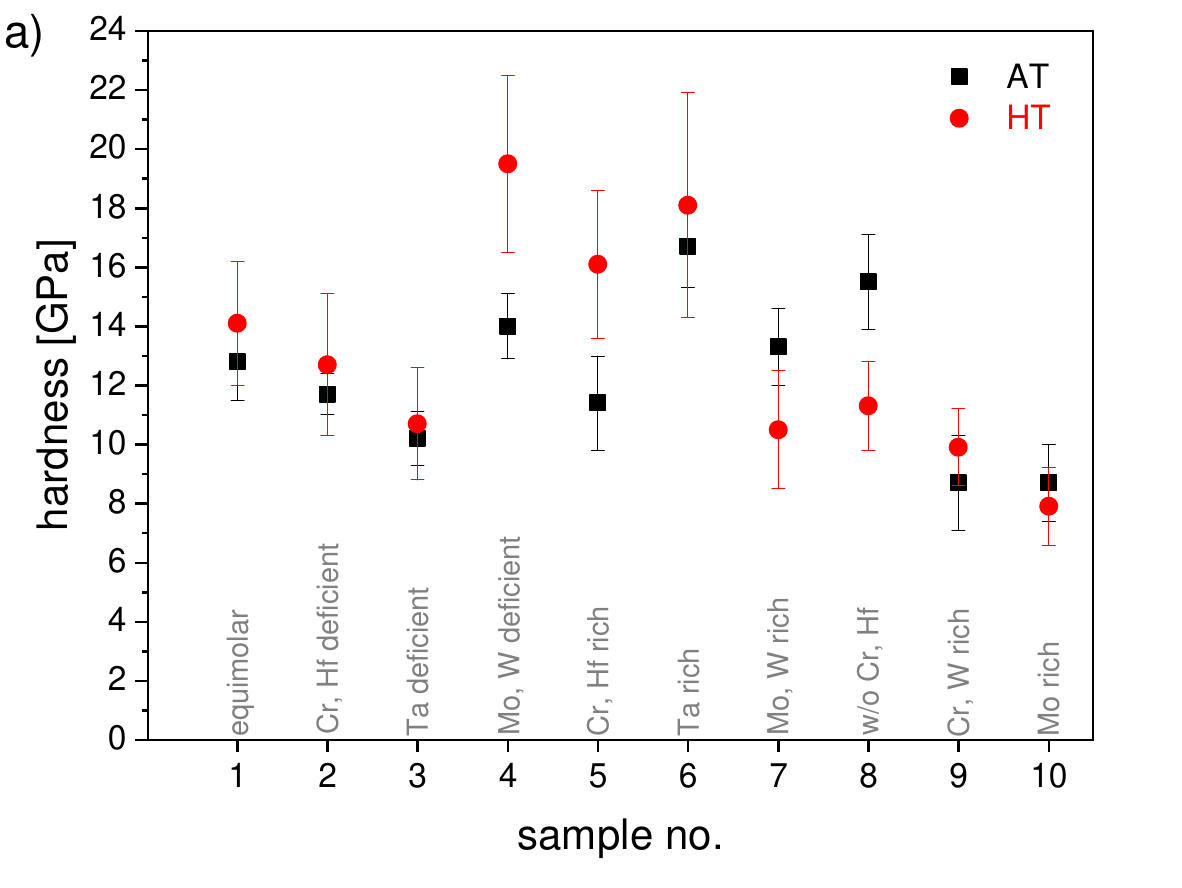}
        \includegraphics[width=.49\textwidth]{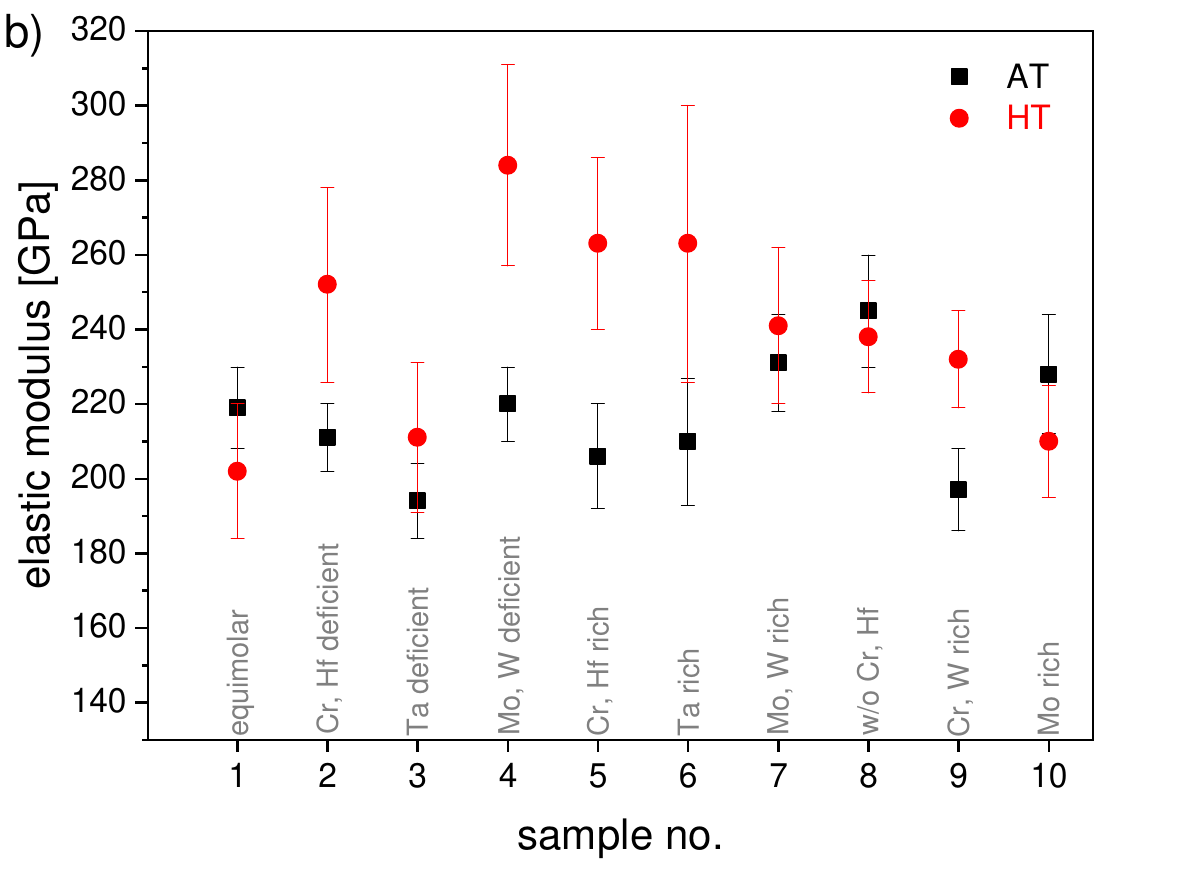}\\
        \includegraphics[width=.49\textwidth]{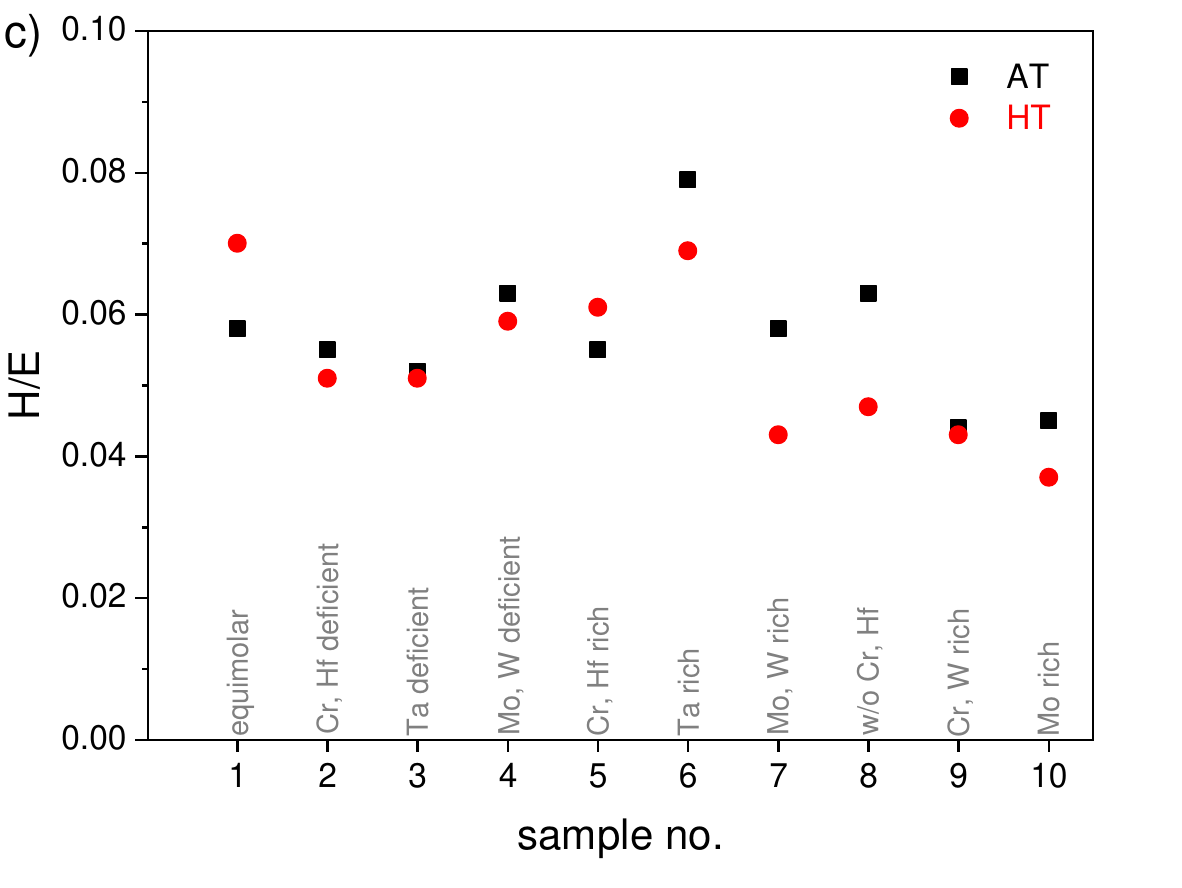}
        \includegraphics[width=.49\textwidth]{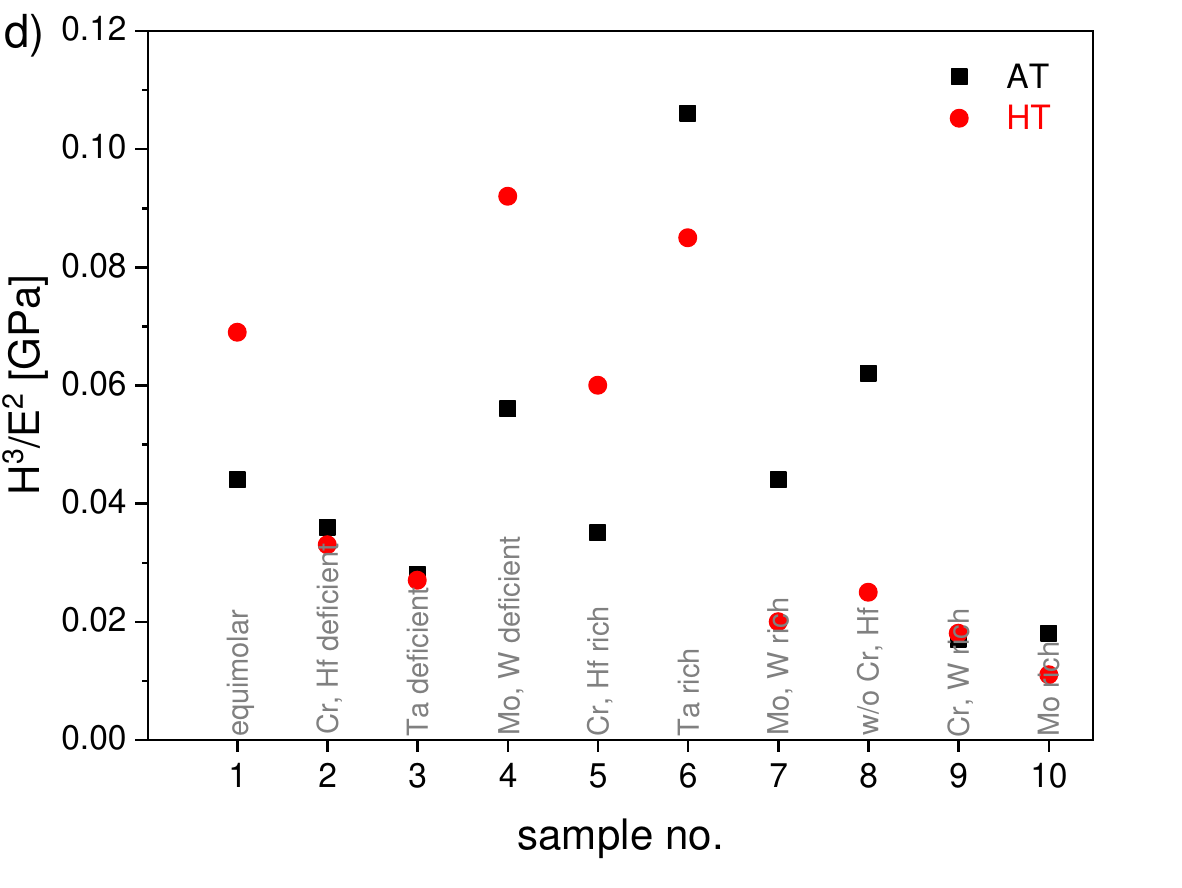}\\
        \caption{a) Hardness (H), b) effective elastic modulus (E), c) H/E, d) H$^3$/E$^2$ ratios of the deposited coatings.}
        \label{mechprop1}
\end{figure}

The mechanical properties of the deposited coatings are shown in Figure~\ref{mechprop1}, with hardness values plotted in Figure~\ref{mechprop1}a. The measured hardness spans a range of approximately 8--20\,GPa. While trends in the AT series are relatively weak, they become more apparent in the HT series. In general, the HT coatings exhibit higher hardness than their AT counterparts, which can be partly attributed to the denser microstructure observed via TEM analysis, and partly to vacancy-induced hardening \cite{KINDLUND2019137479}.
A clear trend emerges regarding the influence of individual elements: molybdenum, tungsten, and chromium tend to reduce hardness, whereas tantalum and hafnium contribute positively. This behaviour may be explained by the known tendency of molybdenum and tungsten nitrides to crystallise in hexagonal rather than cubic phases \cite{hones2003structural}, potentially destabilising the desired fcc structure. This effect is particularly evident when comparing the hardness of samples 7 and 8 in both the AT and HT series. For these samples, increasing the deposition temperature—thereby enhancing diffusion and crystallisation of Mo, W, and Cr—resulted in reduced hardness.
Additionally, chromium nitride itself is known to exhibit relatively low hardness compared to other cubic binary nitrides in the studied system \cite{hones2003structural,HU2015141,BERNOULLI2013157}, further supporting its detrimental effect on the mechanical performance of these coatings.

In our study, the highest hardness values in the high-entropy nitride (HEN) coatings were achieved through a combination of large grain size and a dominant (200) texture. This finding contrasts with the conventional expectation of grain refinement-induced hardening, as described by the Hall–Petch effect, which predicts maximum hardness for coatings with smaller grains, typically around 20\,nm \cite{hall1951deformation}. 
It also deviates from the commonly observed behaviour in ``classical'' low-entropy Ti-based nitrides, where a strong (111) texture—owing to its highest planar atomic density—is generally associated with maximum hardness \cite{MA2004184}. These observations highlight the inherent complexity of high-entropy materials, where conventional structure–property relationships may not apply straightforwardly.


The effective elastic modulus of the coatings is shown in Figure~\ref{mechprop1}b. Its behaviour closely mirrors that of the hardness. Specifically, tantalum and hafnium contribute to an increase in the effective elastic modulus, while tungsten, molybdenum, and chromium tend to reduce it. 
This trend is consistent with the reported Young’s moduli of the corresponding near-stoichiometric binary nitrides. Hafnium nitride (HfN) and tantalum nitride (TaN) exhibit relatively high values, typically in the range of $\sim$350--390\,GPa \cite{HU2015141,TOROK198737} for HfN and $\sim$350--450\,GPa \cite{dastan2022influence} for TaN. In contrast, the moduli of other binary nitrides in the system are lower—approximately 200\,GPa for CrN, $\sim$300\,GPa for fcc substoichiometric WN, and $\sim$310\,GPa for fcc substoichiometric MoN \cite{hones2003structural,klimashin2016impact}.

To enable a more direct comparison of our results with reference values and other high-entropy systems reported in the literature, it is necessary to estimate the Poisson’s ratio of the deposited coatings. \textit{Ab initio} calculations of Poisson’s ratio were performed for five representative coatings: an equimolar composition with a fully occupied nitrogen sublattice, an equimolar composition with 50\,\% nitrogen vacancies, Cr- and Hf-rich, Ta-rich, Mo- and W-rich, and Mo-rich coatings. The calculated Poisson’s ratios ranged from 0.310 to 0.3425, with a mean value of 0.328~$\pm$~0.005.
Using this estimate, Young’s modulus ($Y$) can be calculated from the effective elastic modulus ($E$) using the relation $Y = E \cdot (1 - \nu^2)$, where $\nu$ is the Poisson’s ratio. For $\nu = 0.328$, Young’s modulus corresponds to approximately 90\,\% of the effective elastic modulus. Based on this relation, the Young’s moduli of the coatings fall within the range of approximately 180--250\,GPa.
Although these values are relatively modest, they are comparable to—or even exceed—the values typically reported for refractory-metal-based high-entropy coatings, which are generally around 200\,GPa \cite{SUN2024130775,sheng2016nano}.

The \textit{ab initio} elastic constants' calculations also allowed evaluating the Cauchy pressure, $CP=C_{12}-C_{44}$, commonly used to estimate inherent metallicity of bonding: with more positive/negative values indicating more ductile/brittle behaviour \cite{Pettifor01041992}. HENs with a fully occupied nitrogen sublattice (see Figure~\ref{Calc}a-1) exhibit $CP\sim$80--115\,GPa, suggesting notably higher ductility compared with typical brittle ceramics like TiN or HfN (with $CP$ of $-30$ and $-5$~GPa, respectively). Nitrogen vacancies, however, strongly influence this parameter. For the equimolar coating, the Cauchy pressure decreases from 105\,GPa (fully occupied nitrogen sublattice) to 60\,GPa when 50\,\% nitrogen vacancies are introduced, which may be attributed to increased lattice stresses associated with high vacancy contents.
From an experimental perspective, ductility can be assessed using the $H/E$ ratio, which reflects the ability of a material to sustain elastic strain before failure \cite{LEYLAND20001}. It has been proposed that dense, void-free coatings exhibiting mild residual compressive stress and elastic recovery $\gtrsim$\,60\,\% can be considered ductile if their $H/E$ ratio is $\gtrsim$\,0.1 \cite{musil}. The $H/E$ ratios of the deposited coatings are shown in Figure~\ref{mechprop1}c. Most coatings display values in the range of 0.5--0.8. The presence of molybdenum, tungsten, or chromium reduces the $H/E$ ratio, whereas the coatings with the highest hardness (samples 4--6) also exhibit the highest $H/E$ ratios, suggesting moderate ductility.
While no clear influence of deposition temperature on the $H/E$ ratio was observed, this descriptor should be interpreted with care, particularly given the complex stress states within the films.
The $H^3$/$E^2$ ratio, which correlates with resistance to plastic deformation and is often used as a predictor of tribological performance \cite{MAYRHOFER2003725}, was also evaluated. However, no clear trend was observed across the coatings, indicating that this parameter cannot reliably predict their tribological behaviour in this case.

\subsection{Temperature stability analysis}
One sample deposited on silicon substrates was annealed under vacuum at 1000\textdegree C and, one sample at 1200\textdegree C, respectively. An additional sample, deposited on R-plane cut sapphire, was annealed at 1400\textdegree C, after which a small amount of air was introduced to assess the onset of oxidation.

Most of the AT coatings, with the exception of the medium-entropy Mo-Ta-W-N coating AT8, did not withstand the initial annealing at 1000\textdegree C on silicon substrates and delaminated. In contrast, all HT coatings, except HT9 and HT10, withstood this annealing step. Improved adhesion at higher deposition temperatures during magnetron sputtering is a well-known phenomenon, often attributed to adatoms possessing higher diffusion energy, enabling them to migrate to lower-energy sites on the substrate. This promotes better surface coverage by overcoming shadowing effects associated with low-temperature depositions, reduces intrinsic stresses in the coating, and may facilitate the formation of a thin diffusion interlayer at the substrate–coating interface \cite{MURTY1998328,eltoukhy79,Windischmann01011992,WARRES2023139977}. Another possible explanation is the denser microstructure observed in HT coatings via TEM. In contrast, the AT series exhibits column boundaries with elevated oxygen content, which may serve as crack initiation sites, ultimately leading to coating failure.

\subsubsection{Chemical composition}
\begin{figure}[htpb]
        \centering
        \includegraphics[width=.49\textwidth]{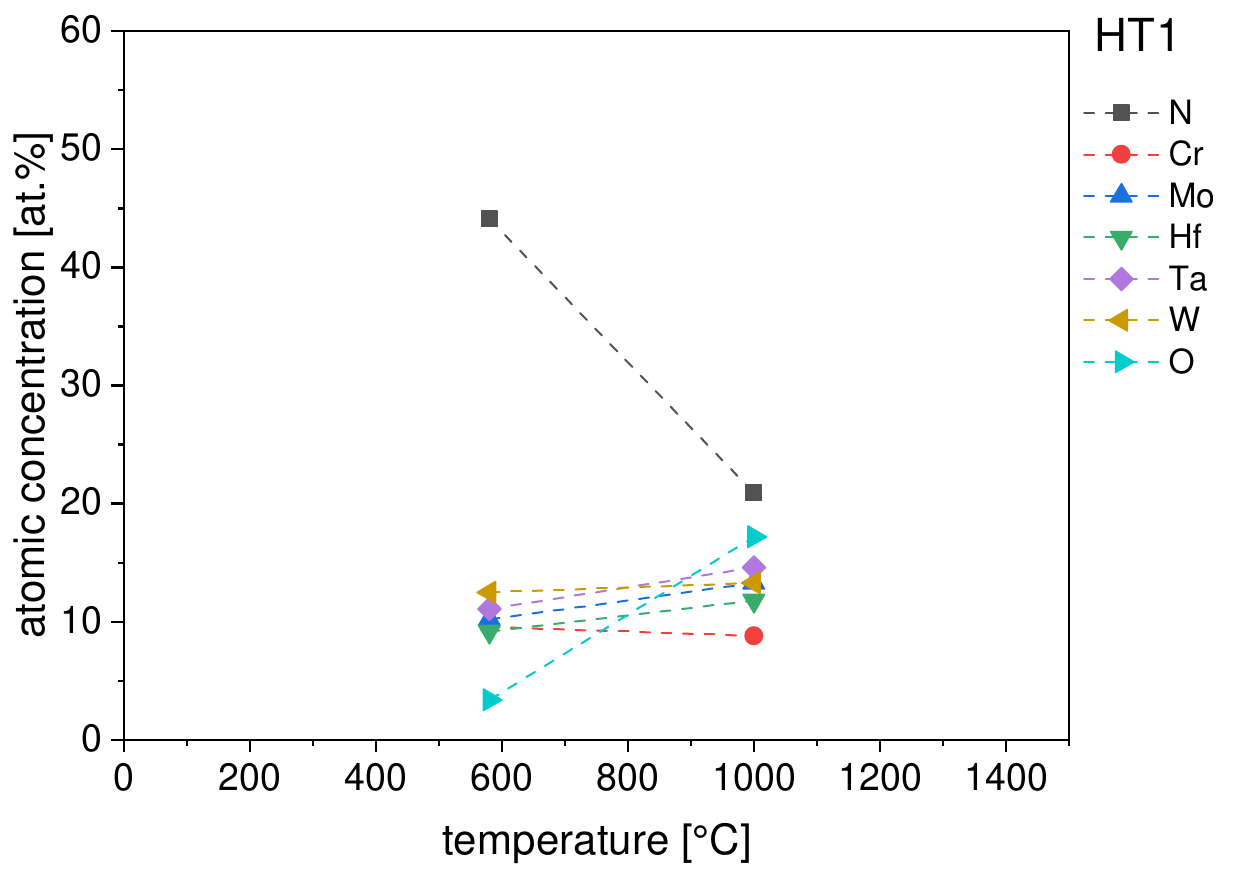}
        \includegraphics[width=.49\textwidth]{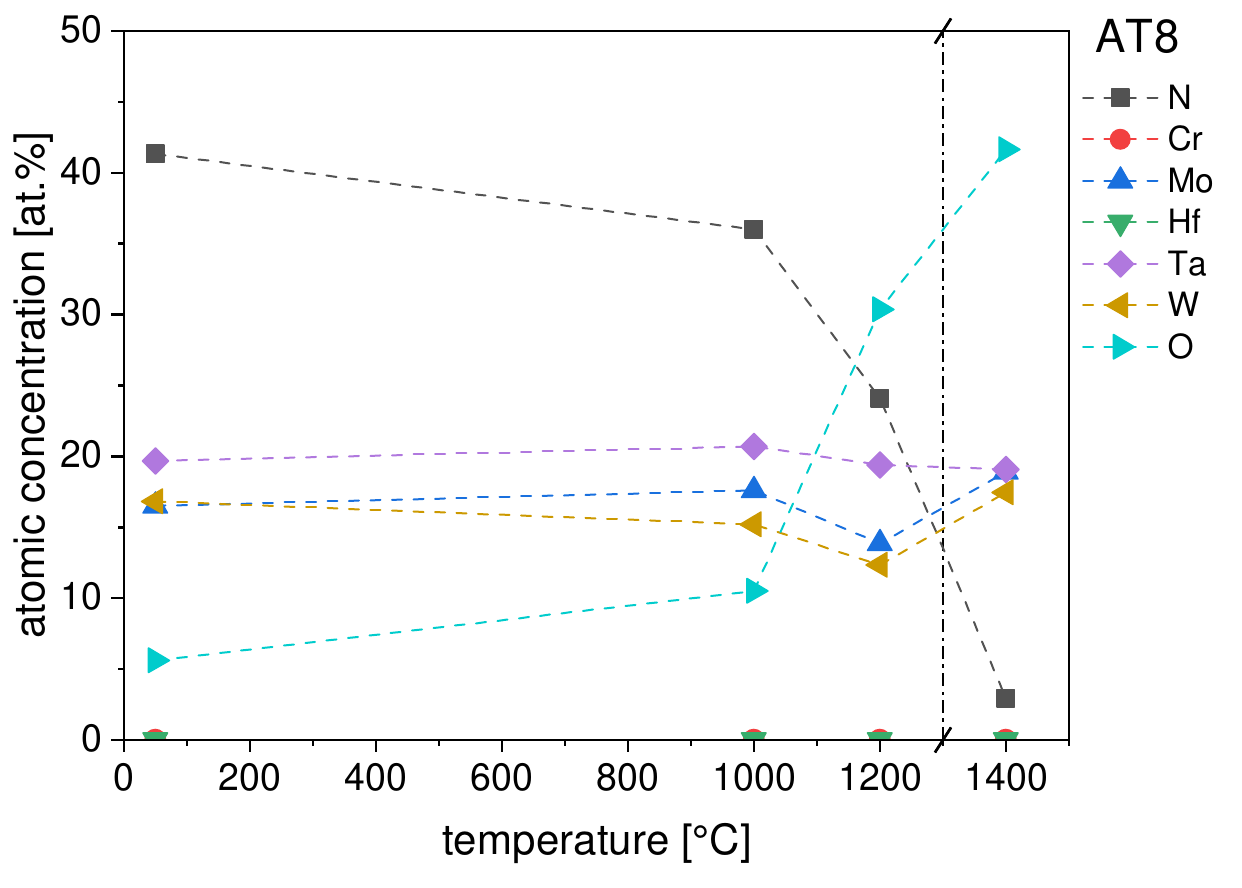}\\
        \includegraphics[width=.49\textwidth]{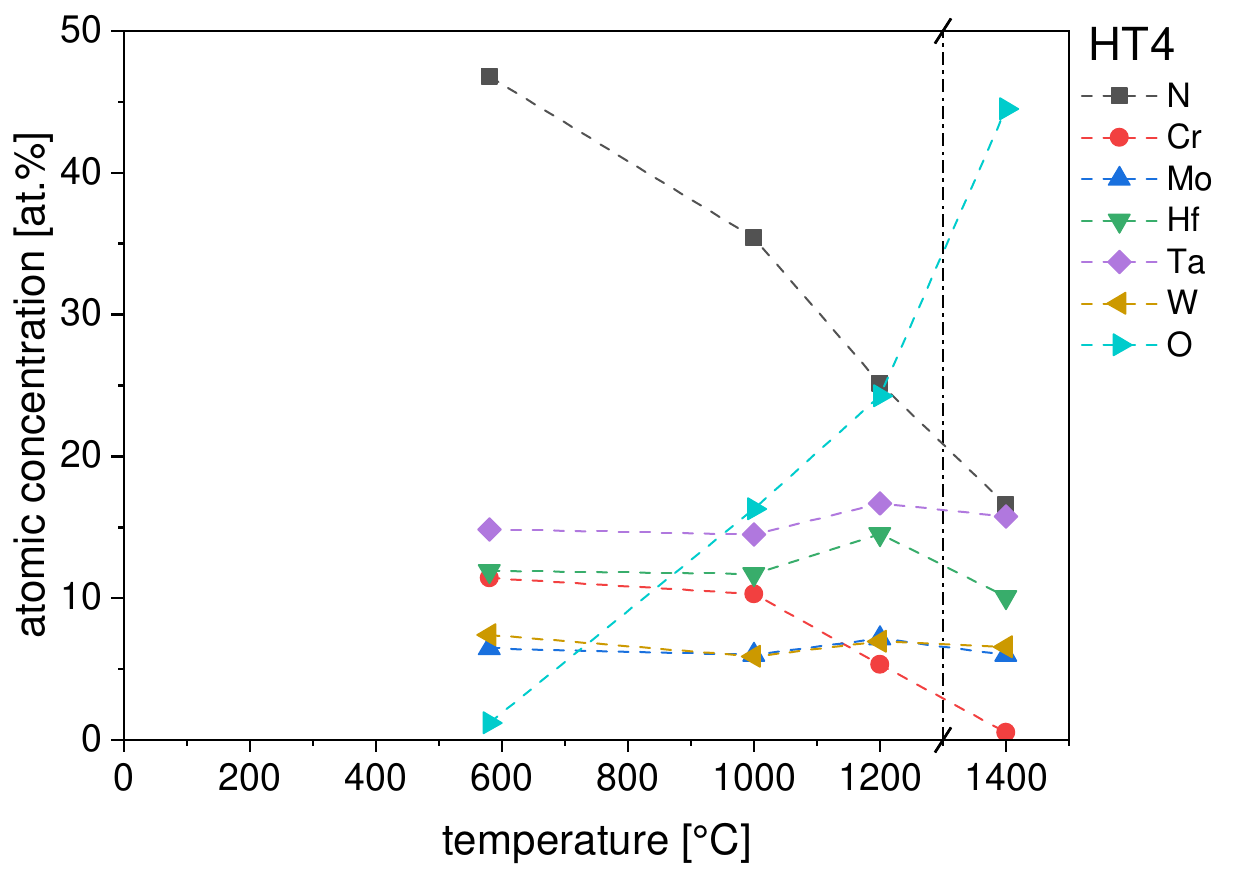}
        \includegraphics[width=.49\textwidth]{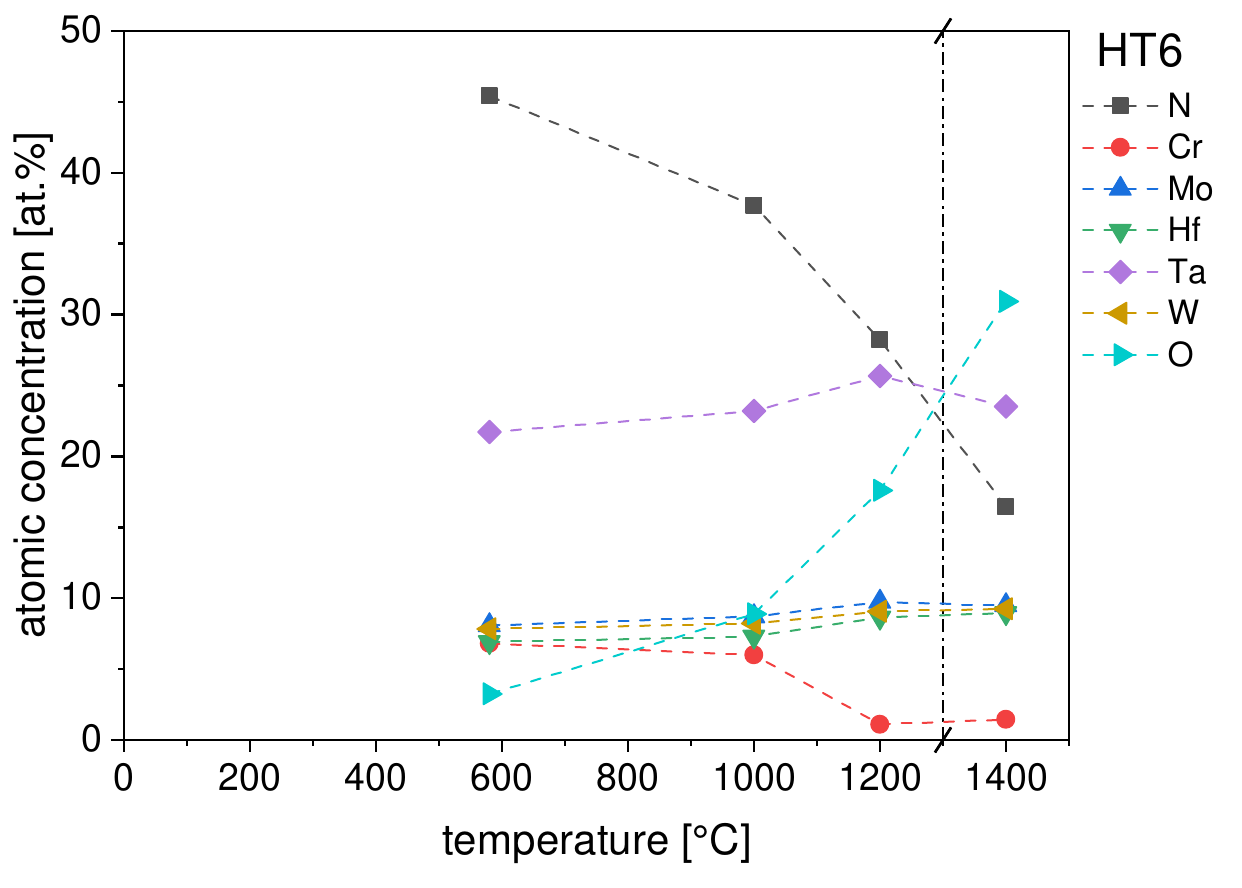}\\
        \includegraphics[width=.49\textwidth]{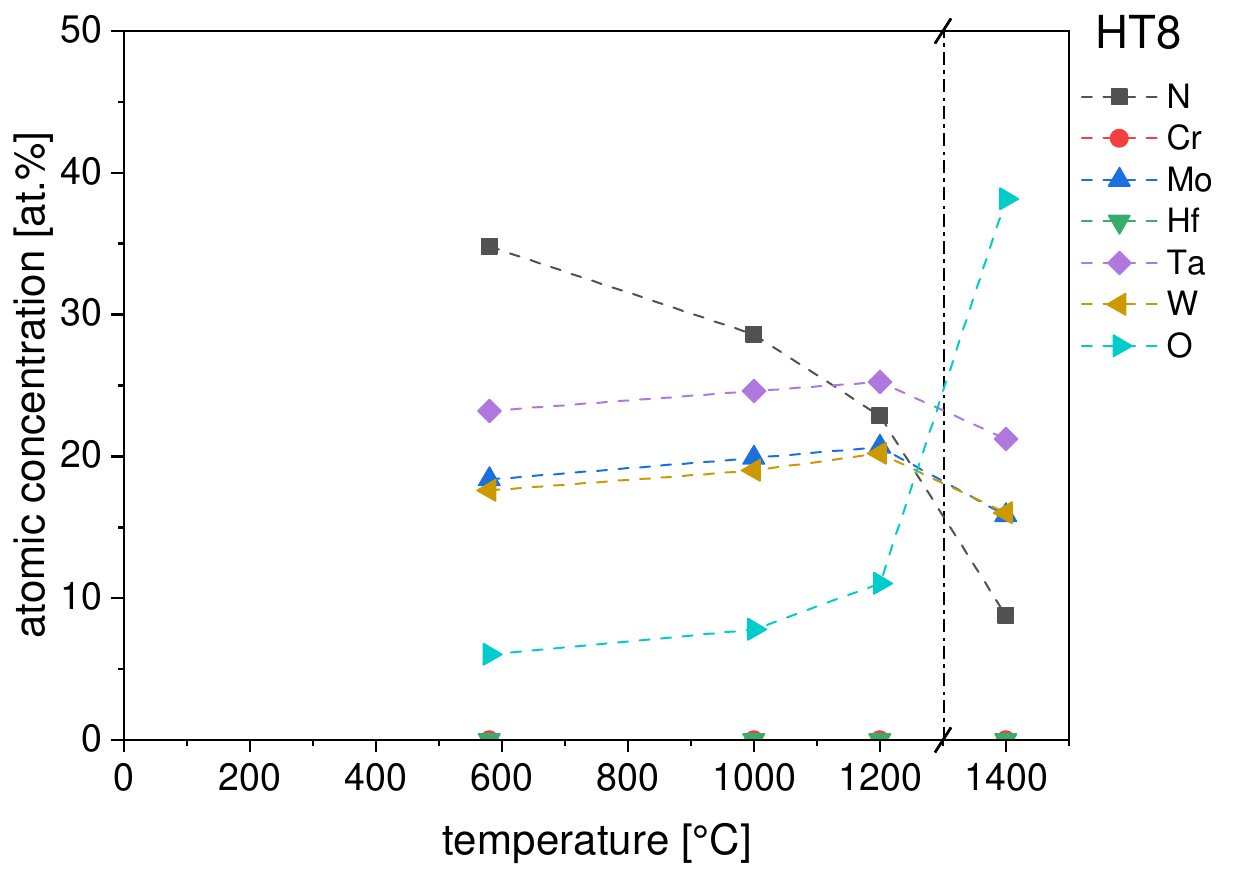}\\
        \caption{The evolution of the chemical composition after annealing of samples HT1 representing coatings that did not withstand annealing at temperatures higher than 1000\textdegree C and samples AT8, HT4, HT6 and HT8. Initial points represent the deposition temperature.}
        \label{heat_chem}
\end{figure}

The evolution of the chemical composition of all coatings that withstood the initial annealing at 1000\textdegree C, for all subsequent heating steps they endured, is presented in Figure S4 in the Supplementary Material. As oxidation was also examined, the oxygen content is included. The coatings can be categorised into two groups. The first group, comprising HT1, HT2, HT3, HT5, and HT7, withstood only the annealing at 1000\textdegree C and delaminated from the silicon substrate at 1200\textdegree C. The second group, consisting of AT8, HT4, HT6, and HT8, withstood all annealing temperatures. The chemical compositions of a representative coating from the first group (HT1) and the coatings that withstood higher temperatures are shown in Figure \ref{heat_chem}.  

All coatings experienced nitrogen loss during annealing. The rate of nitrogen depletion distinguished those that delaminated after 1000\textdegree C from those that endured higher temperatures. Coatings that failed at 1200\textdegree C exhibited a pronounced nitrogen loss: from an as-deposited content of $\sim$40--45~at.\%, their nitrogen content dropped to below 30~at.\% after annealing at 1000\textdegree C, reaching 15--27~at.\%. In contrast, coatings AT8, HT4, and HT6, which withstood all annealing steps, retained $\gtrsim$40~at.\% nitrogen from initial values of $\sim$44--50~at.\% after the first annealing step. Coating HT8, with 37~at.\% nitrogen as-deposited, retained 29~at.\% after annealing at 1000\textdegree C. Thus, coatings with nitrogen losses exceeding 20~at.\% after the first annealing did not withstand the subsequent step at 1200\textdegree C, whereas those with losses below 10~at.\% endured higher temperatures.  

No oxygen was intentionally introduced during these annealing steps; however, the observed oxygen uptake correlated with nitrogen loss. This fact, together with no microstructural changes observed by XRD as discussed in the next section, suggests that the increased oxygen content mostly arose from post-annealing atmospheric contamination due to enhanced microscopic porosity. Another notable effect was the loss of chromium at temperatures above 1000\textdegree C, likely linked to the lower thermal stability of chromium nitrides \cite{Widenmeyer2014}.

When the annealing temperature was increased to 1400\textdegree C and air was introduced, the nitrogen content in the samples decreased further, accompanied by a corresponding increase in oxygen content. At this stage, in addition to oxygen physisorption associated with porosity, chemical reactions and phase transformations are also expected to occur. Further loss of chromium in this case may be attributed to the formation of a volatile chromium oxide, such as CrO$_3$, and its consecutive release from the coating.

\subsubsection{Morphology and microstructure}
The morphological evolution of the coatings is illustrated in Figure \ref{heat_sem} for samples AT8 and HT8, with additional examples (HT1, HT4, and HT6) provided in the Supplementary Material (Figures S5--S7).  
Sample AT8 exhibits a dense microstructure in both the as-deposited state and after annealing at 1000\textdegree C, corresponding to a low oxygen concentration in each case. After annealing at 1200\textdegree C, the structure, particularly at the surface, becomes noticeably coarser, accompanied by a significant increase in oxygen content. Annealing at 1400\textdegree C with air introduction leads to further coarsening and pronounced cracking.  
Sample HT8 maintains an essentially unchanged microstructure after annealing at 1000\textdegree C and 1200\textdegree C, with oxygen content remaining low. A marked increase in oxygen concentration occurs only after air is introduced at 1400\textdegree C, coinciding with the onset of cracking.  
Samples HT4 and HT6 display a gradual coarsening of the microstructure throughout the annealing process, accompanied by steadily increasing oxygen levels. These results suggest that the higher oxygen content originates from post-deposition physisorption, facilitated by increased coating porosity across multiple length scales.

 \begin{figure}[htpb]
        \centering
        \includegraphics[width=.48\textwidth]{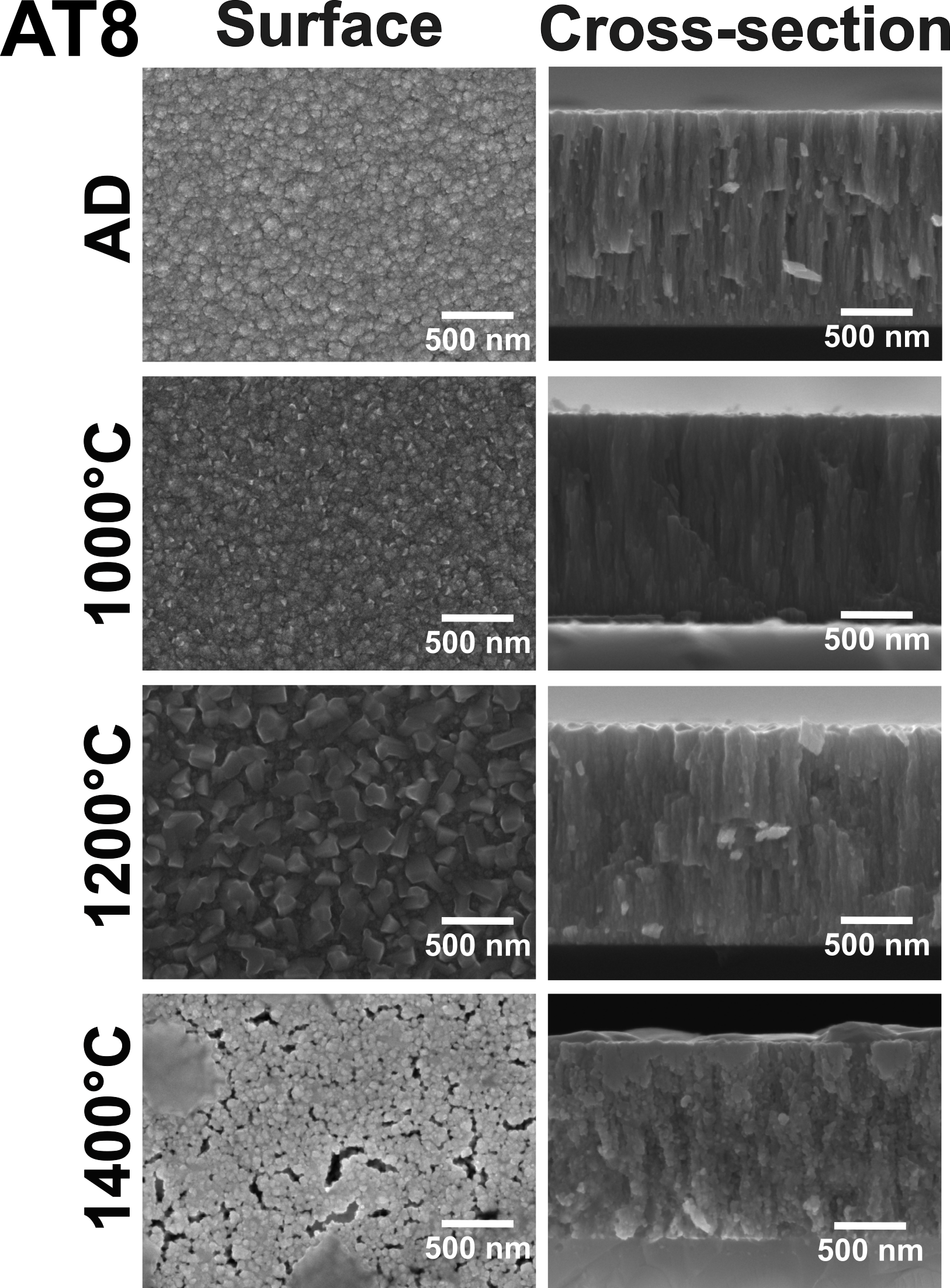}
        \includegraphics[width=.48\textwidth]{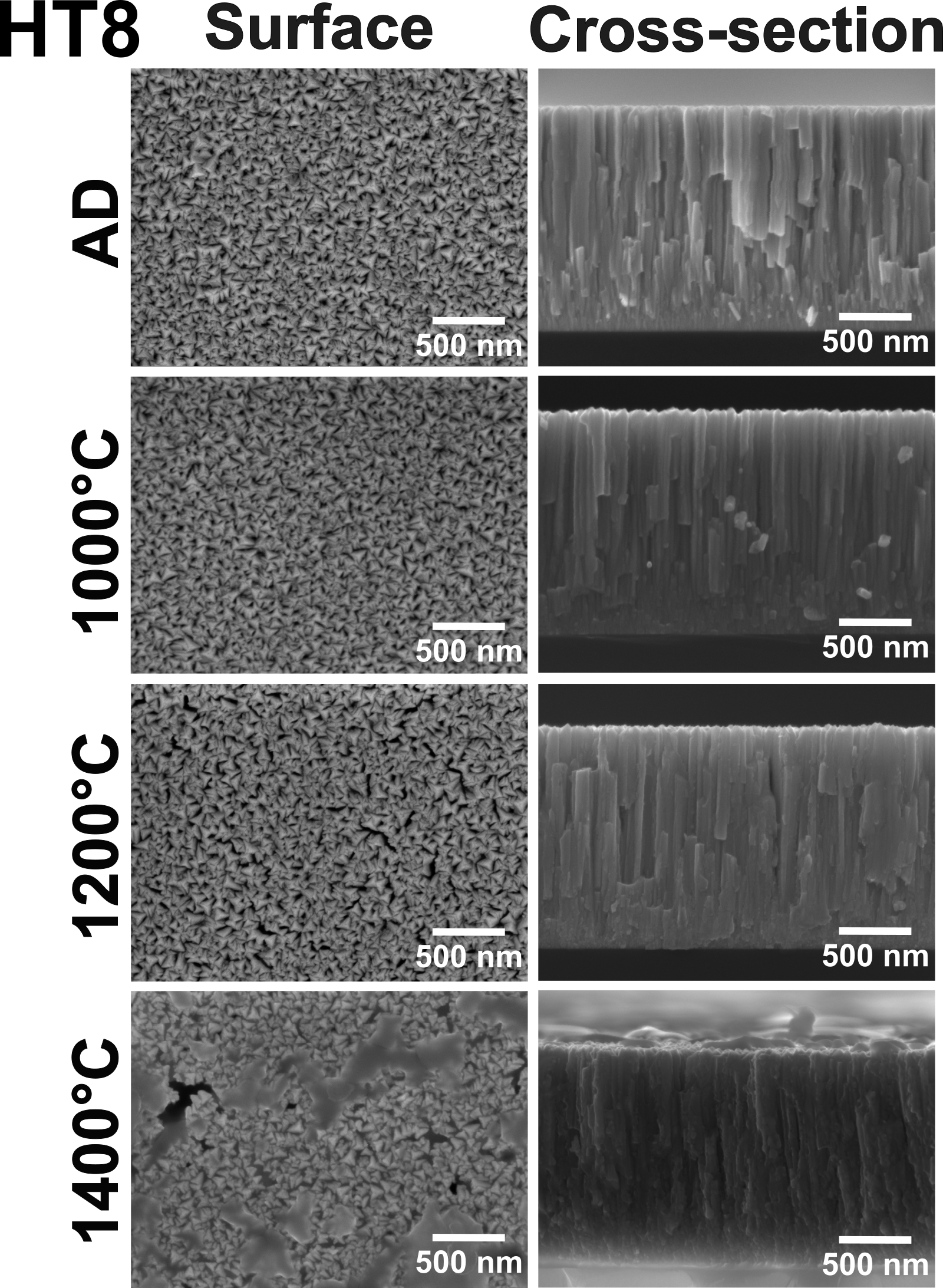}\\
        \caption{The evolution of the morphology of samples AT8 and HT8 after annealing.}
        \label{heat_sem}
\end{figure}

\begin{figure}[htpb]
        \centering
        \includegraphics[width=.49\textwidth]{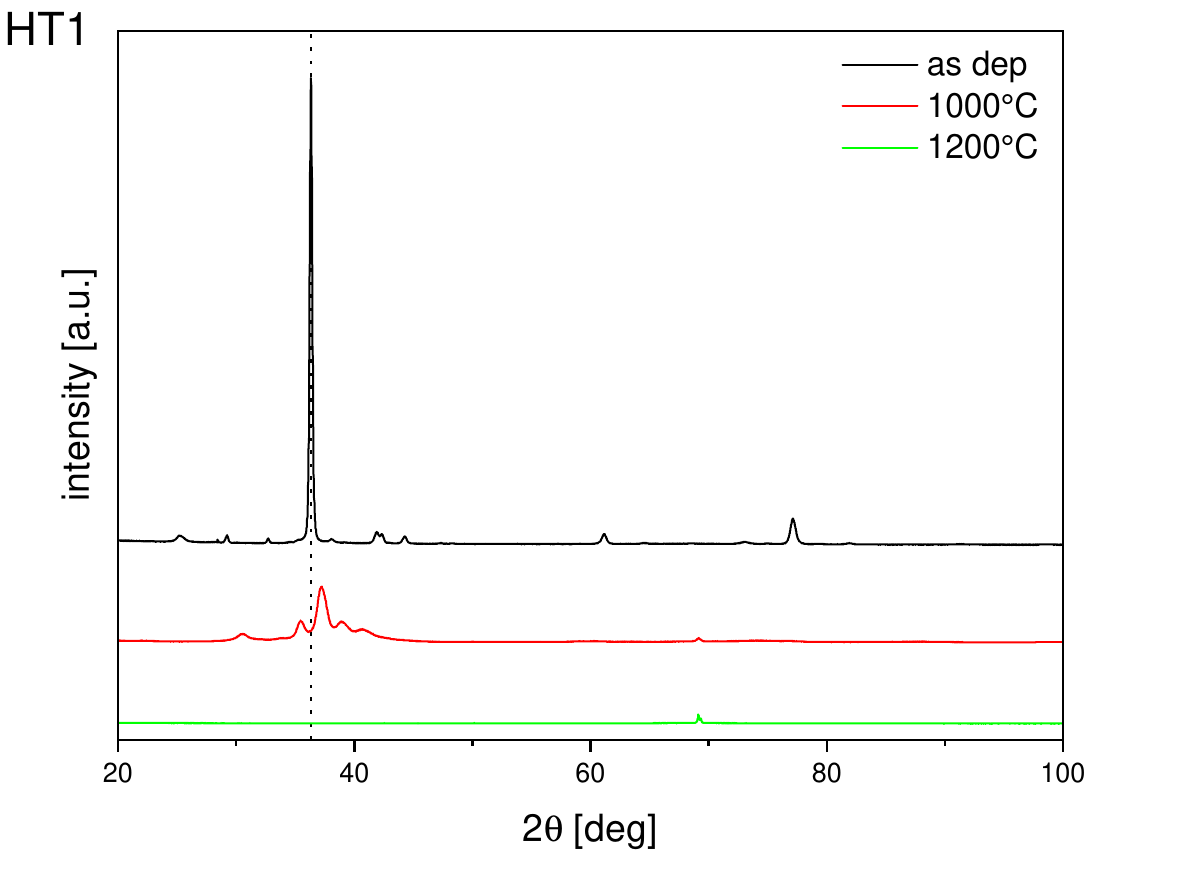}
        \includegraphics[width=.49\textwidth]{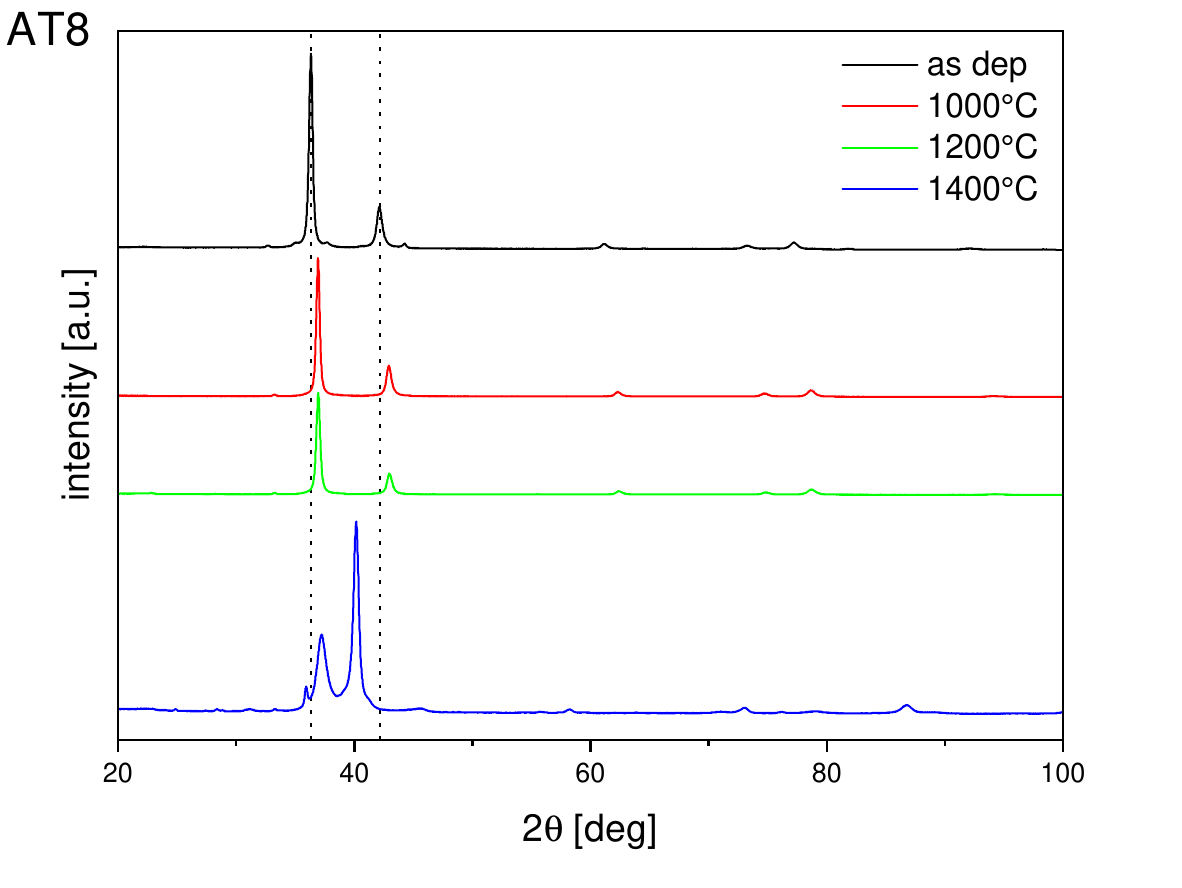}\\
        \includegraphics[width=.49\textwidth]{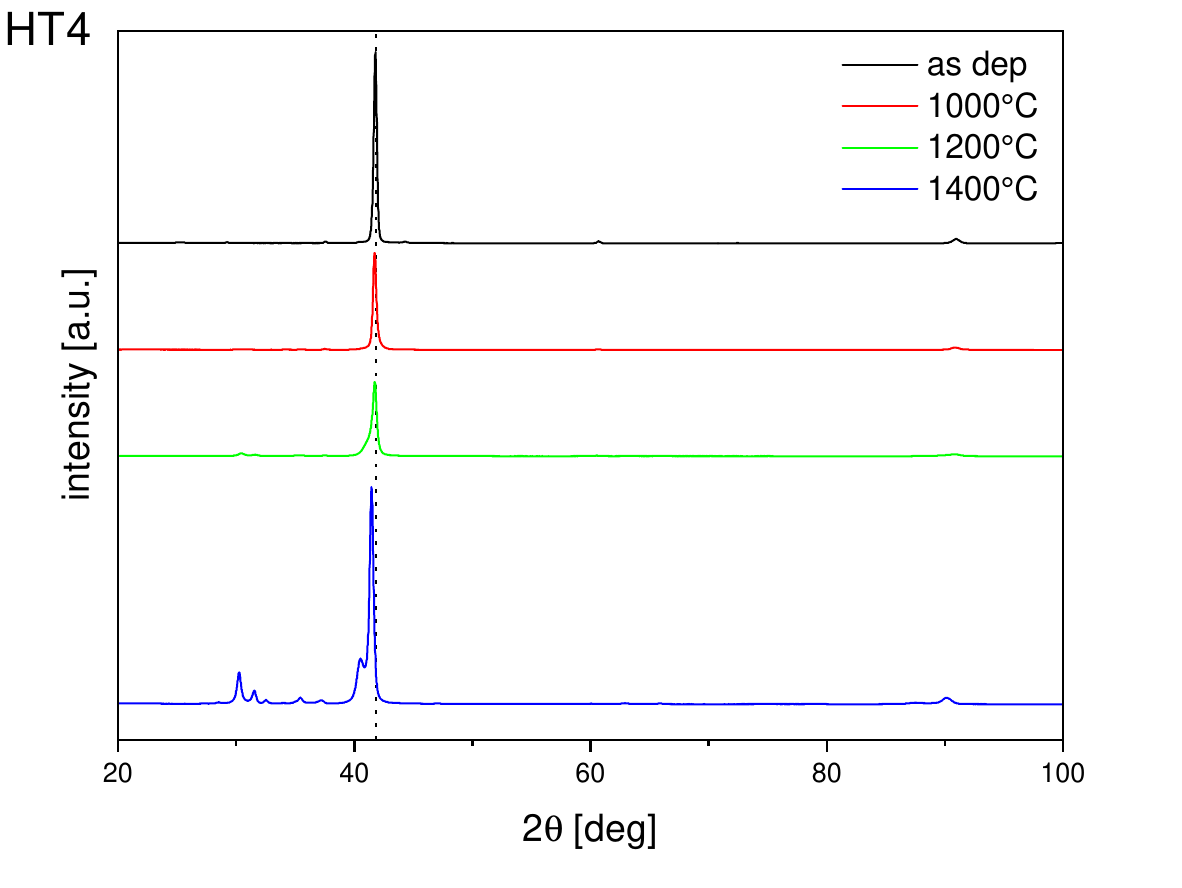}
        \includegraphics[width=.49\textwidth]{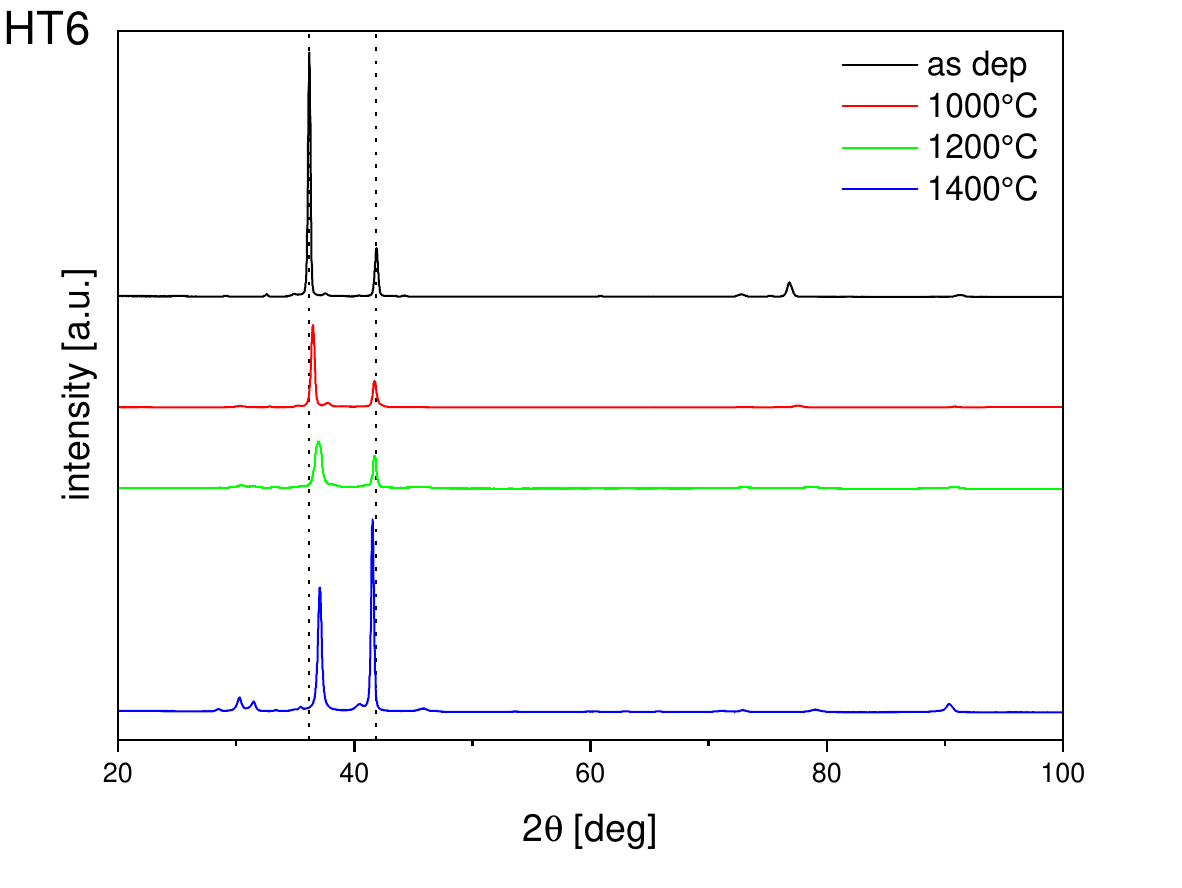}\\
        \includegraphics[width=.49\textwidth]{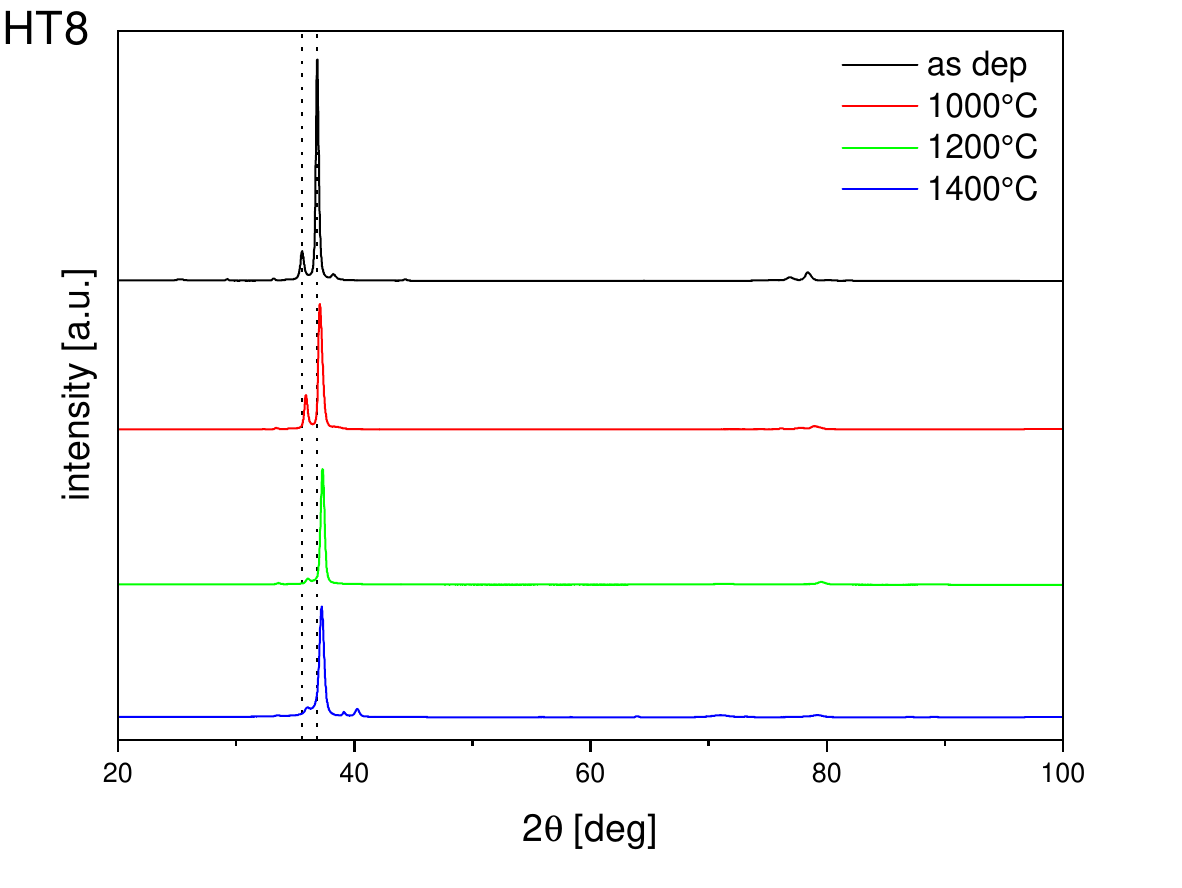}\\
        \caption{X-ray diffractograms of samples HT1 representing coatings that did not withstand annealing at temperatures higher than 1000\textdegree C and samples AT8, HT4, HT6 and HT8. Initial points represent the deposition temperature.}
        \label{heat_xrd}
\end{figure}

The X-ray diffractograms of all coatings that withstood the initial annealing at 1000\textdegree C, for all subsequent heating steps they endured, are presented in Figure S8 in the Supplementary Material. Diffractograms for a representative coating from the group that failed above 1000\textdegree C (HT1) and for all coatings that withstood higher temperatures are shown in Figure \ref{heat_xrd}.  

Sample HT1 exhibits a markedly lower diffraction peak intensity after the first annealing step, along with the appearance of additional peaks. These changes indicate that nitrogen loss at 1000\textdegree C is coupled with crystalline structure decomposition. After annealing at 1200\textdegree C, no crystalline reflections from the coating are detected, and SEM observations reveal that only traces of the coating remain.  

The Mo-Ta-W-N medium-entropy coating AT8 shows no significant changes in crystalline structure up to 1200\textdegree C. At 1400\textdegree C with air introduction, nearly all nitrogen is desorbed and replaced by oxygen, and the observed peaks correspond to a mixture of cubic metallic phases and oxides. The Mo- and W-deficient coating HT4 demonstrates excellent thermal and oxidation stability, with only a satellite peak and a few smaller reflections at $\sim$30--35\textdegree\, indicating the onset of phase transformation after oxidation.  

The Ta-rich coating HT6 displays two strong peaks in all conditions, including after oxidation. In the as-deposited state and after annealing at 1000\textdegree C, their positions are consistent with a single fcc phase. At higher temperatures, however, the peaks shift in opposite directions: the $\sim$36\textdegree\, peak moves to higher angles, indicating a reduced lattice parameter, while the $\sim$42\textdegree\, peak shifts to lower angles, indicating an increase in lattice parameter. This behaviour suggests either the formation of a two-phase structure due to changes (mainly oxygen uptake and chromium loss) in the chemical composition or stress relaxation, with both compressive and tensile stresses present in different regions of the as-deposited coating. The shifts become more pronounced after oxidation.  

Coating HT8 exhibits a smaller peak at 35.7\textdegree\, and a dominant peak at $\sim$37\textdegree. The latter may arise from diffraction of the W$_{\textrm{L}\alpha}$ line from the X-ray source on the same phase or from a secondary phase. Both peaks remain unchanged after annealing at 1000\textdegree C. Following annealing at 1200\textdegree C, the smaller peak disappears and the main peak at $\sim$37\textdegree\, broadens, consistent with either the loss of a secondary phase or the disappearance of the W$_{\textrm{L}\alpha}$ line diffraction due to grain refinement. Oxidation testing at 1400\textdegree C produces no significant changes in the diffractogram.

\begin{figure}[htpb]
        \centering
        \includegraphics[width=.6\textwidth]{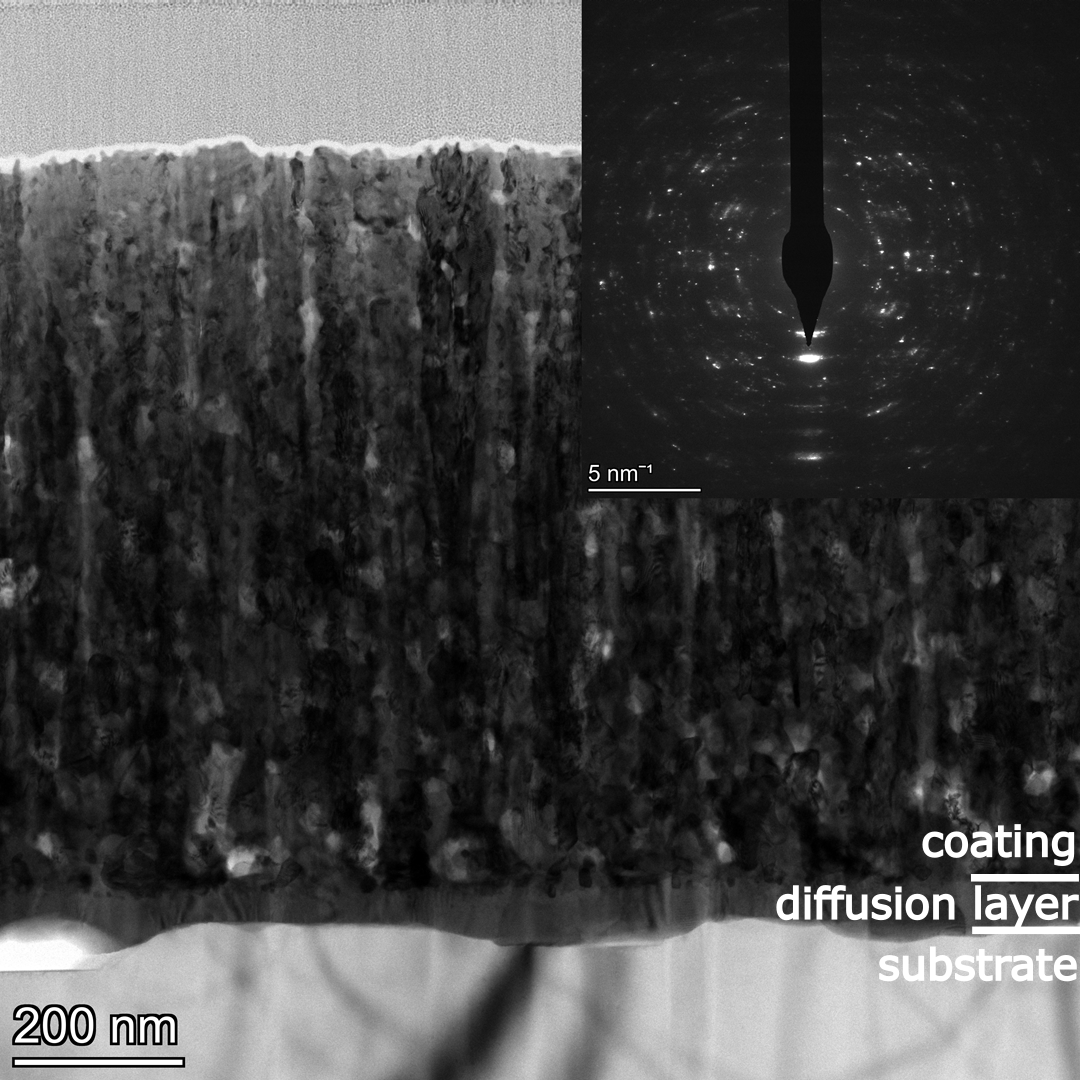}\\
        \caption{TEM micrograph of coating HT4 after oxidation testing at 1400\textdegree C. SAED pattern is in the inset.}
        \label{heat_tem1}
\end{figure}

\begin{figure}[htpb]
        \centering
        \includegraphics[width=1\textwidth]{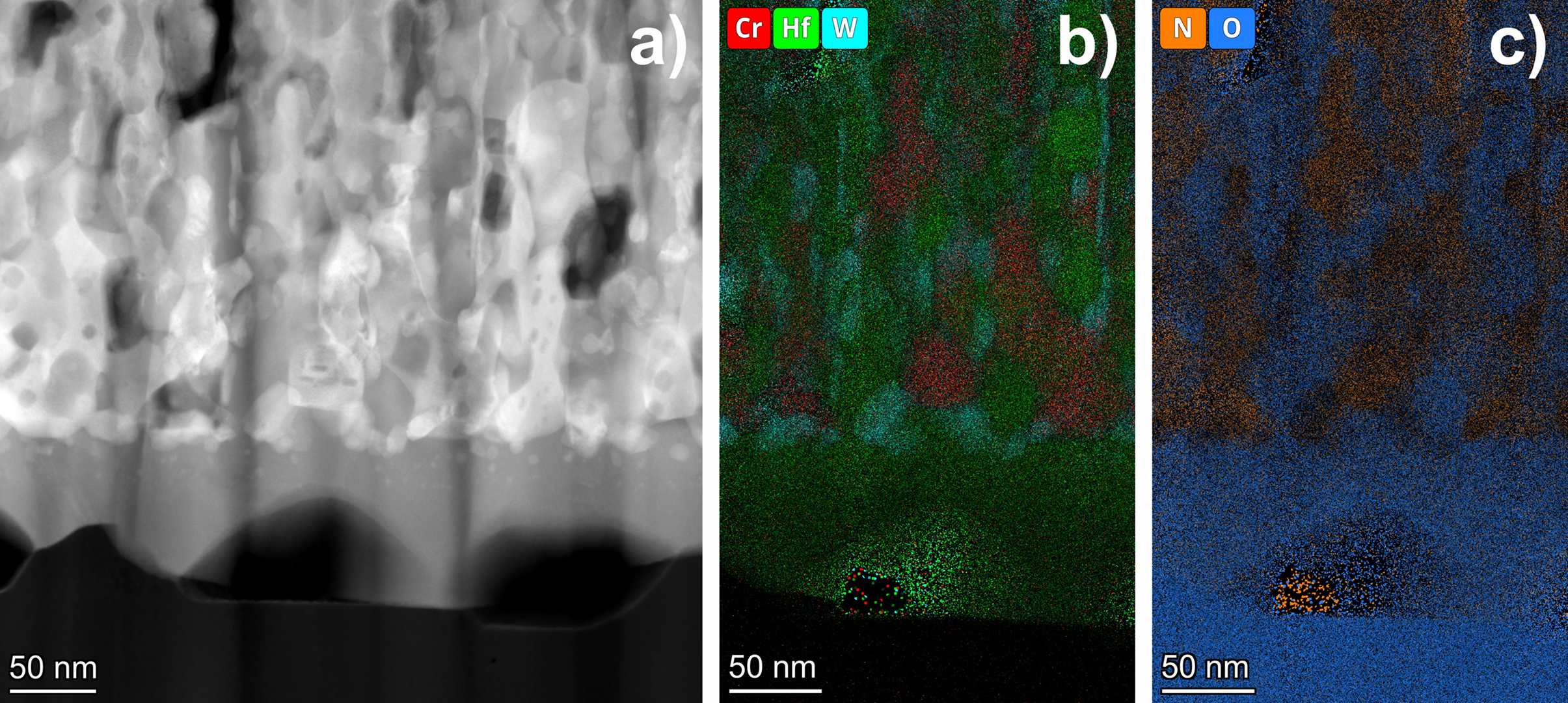}\\
        \caption{STEM HAADF micrograph and EDX elemental maps of the bottom part of sample HT4 after oxidation tests.}
        \label{heat_tem2}
\end{figure}

To examine the changes in the coating at the onset of oxidation in more detail, sample HT4 annealed and oxidised at 1400\textdegree C was further characterised by TEM. The bright-field image of the entire sample is shown in Figure \ref{heat_tem1}, with the corresponding SAED pattern in the inset. The microstructure shows a clear transformation, evident from both the film morphology—where $\sim$50~nm globular grains are observed—and from changes in the SAED pattern. The corresponding elemental distribution maps are presented in \ref{heat_tem2}. 

These observations indicate the presence of three distinct phases: (i) an fcc transition-metal nitride (Mo, Ta, Cr) with a lattice parameter similar to that of the unoxidised coating, corresponding to the main XRD peak -- the distribution of Cr (coincides with Mo and Ta distribution) in Figure \ref{heat_tem2}b indicates the location of the HEA nitride phase; (ii) a metallic bcc tungsten phase, as tungsten does not correlate with either oxygen or nitrogen (see Figures \ref{heat_tem2}b and \ref{heat_tem2}c); and (iii) an HfO$_2$ phase, as the distribution of hafnium correlates with oxygen, while it is independent of the other metals. All these phases are also confirmed by the SAED pattern analysis, which also revealed that the hafnium dioxide phase contains both the monoclinic (PCPDF card 78-0050) and the orthorhombic (PCPDF card 83-0808) allotropes.  

The coating also interacts with the sapphire substrate at the interface, as shown in the lower region of \ref{heat_tem2}. Hafnium and tantalum diffuse into the substrate, forming a $\sim$50--100~nm thick transition layer, with $\sim$50--100~nm voids located between this layer and the substrate. EDX analysis of the transition layer reveals an approximate composition of 10~at.\% Hf, 10~at.\% Ta, 15~at.\% Al, and 60~at.\% O. FFT analysis of HR image of the transition diffusion layer (not shown) indicates a large unit cell with nearly perpendicular axes, as indicated by the 88.2\textdegree\ angle between the 536 pm and 500 pm reflections.  

Based on these parameters, the transition layer may be tentatively identified as orthorhombic based on the Hf$_6$Ta$_2$O$_{17}$ phase (ICSD card 403368) with reference lattice parameters $a = 4.83$~\AA, $b = 4.94$~\AA, and $c = 5.26$~\AA, where the $b$ and $c$ values closely match the measured spacings of 500 and 536~pm, respectively. Aluminium substitution on the metallic sublattice could account for the observed deviations in cell parameters and slight distortions of the unit cell angles. Nevertheless, phase identification in such a complex system cannot be considered unambiguous.

\begin{figure}[htpb]
        \centering
        \includegraphics[width=.53\textwidth]{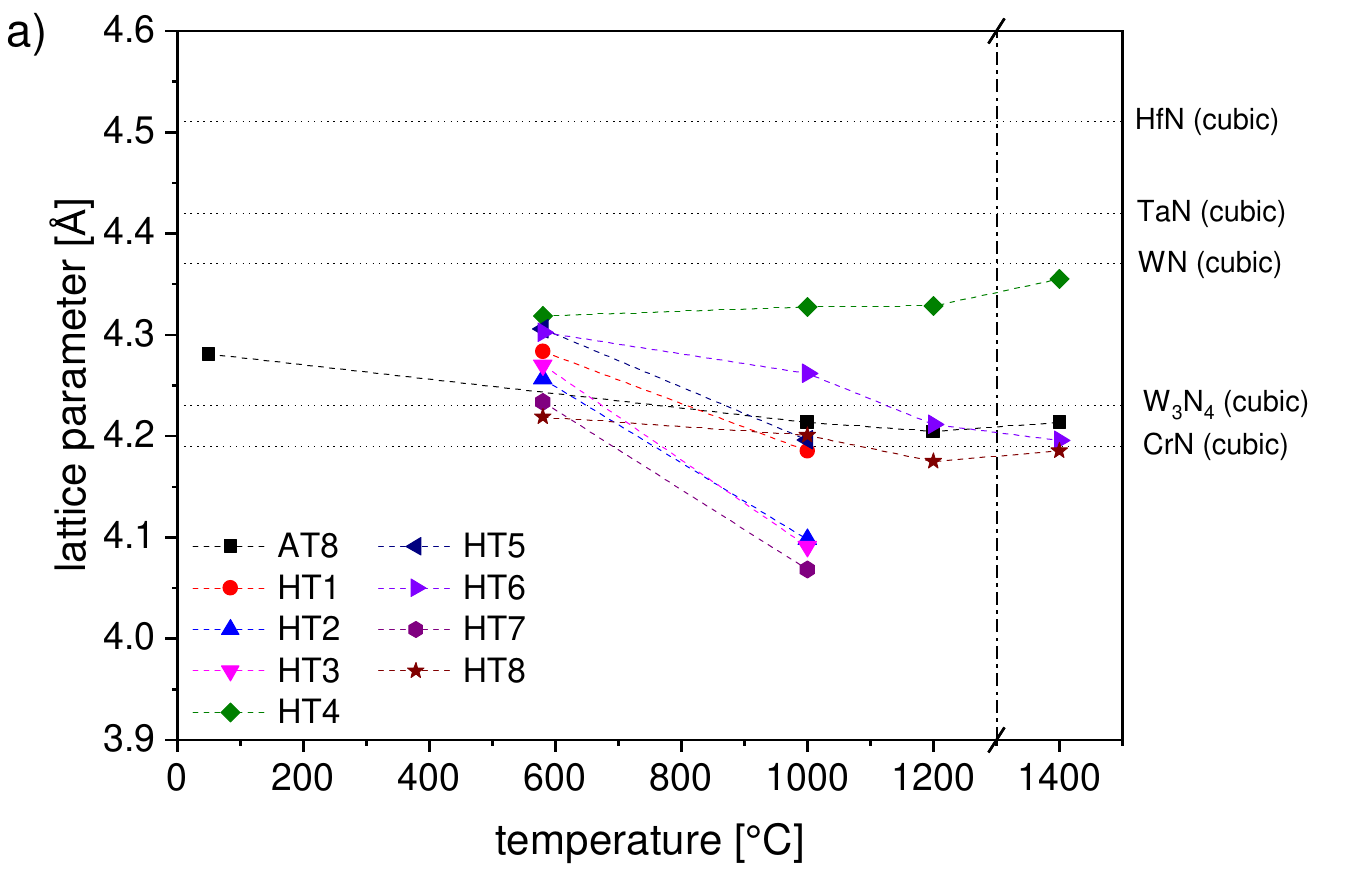}
        \includegraphics[width=.46\textwidth]{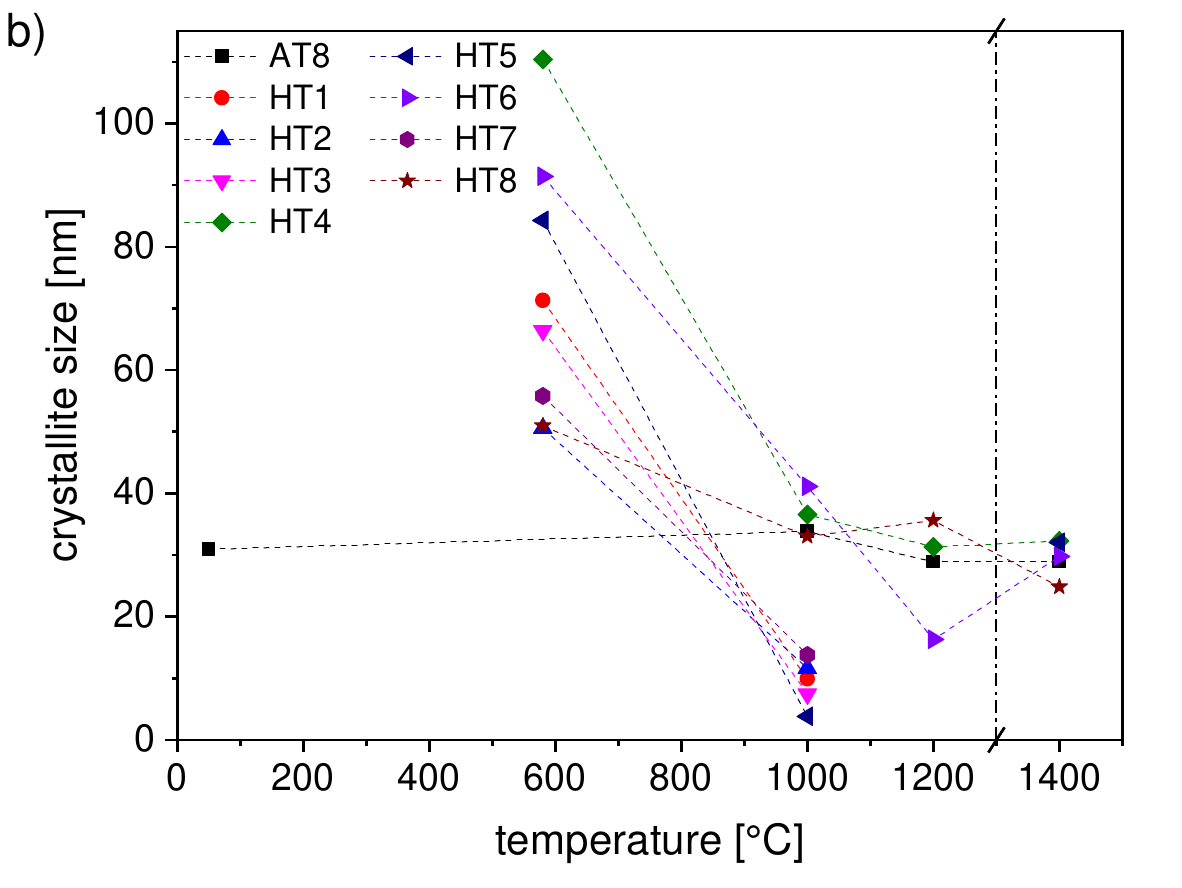}\\
        \caption{a) Lattice parameters and b) crystallite sizes of annealed coatings. Initial points represent the deposition temperature.}
        \label{heat_structure}
\end{figure}

The lattice parameters and crystallite sizes of the coatings, determined from XRD measurements of the main fcc-nitride peak, are shown in Figure \ref{heat_structure}a. Reference values shown as dashed horizontal lines were taken from the Materials Project database \cite{matproj}. All coatings, except HT4, exhibit a decrease in lattice parameter with increasing temperature, consistent with nitrogen release. Coatings that failed above 1000\textdegree C experienced the highest nitrogen loss rates and correspondingly the largest reductions in lattice parameter. In contrast, HT4 demonstrated the strongest nitrogen retention across all annealing temperatures, with its lattice parameter remaining largely unaffected—possibly due to the accommodation of lattice strain, which offsets the relatively low concentration of nitrogen vacancies. After oxidation testing, a slight lattice parameter increase is observed in all coatings, suggesting that oxygen incorporation into the lattice more than compensates for nitrogen loss.  

The crystallite sizes are presented in Figure \ref{heat_structure}b. The only sample deposited at ambient temperature, AT8, had the smallest crystallites in the as-deposited state, and their size was not significantly altered by annealing, even after oxidising at 1400\textdegree C. All coatings deposited at elevated temperatures exhibited much coarser crystallites initially. After annealing at 1000\textdegree C, a pronounced crystallite size reduction was observed. Coatings that failed above 1000\textdegree C showed the greatest decrease, with crystallite sizes determined from the XRD falling below 20\,nm, possibly due to accumulation of a high concentration of dislocations or planar defects. In contrast, coatings that withstood higher temperatures retained crystallite sizes above 30\,nm after the 1000\textdegree C anneal. Further annealing caused only minor additional refinement.

We can conclude that no clear correlation was observed between the thermal stability and the mixing entropy of the deposited coatings. Notably, the nearly equimolar nitride coating HT1, which exhibits the highest configurational entropy on the metallic sublattice, degraded during annealing at 1200\textdegree C. In contrast, the medium-entropy coating HT8, which possesses the lowest mixing entropy among the studied compositions, remained stable during both high-temperature annealing and oxidation testing. This trend persisted across the full compositional range and suggests that, within the investigated system, configurational entropy alone does not serve as a reliable predictor of high-temperature phase stability.

This observation brings to light the interplay between entropic and enthalpic stabilisation mechanisms in multicomponent systems. High-entropy materials are theoretically stabilised by the entropic contribution to the Gibbs free energy, which can suppress the formation of competing phases.  However, our results indicate that enthalpic factors, such as the formation enthalpies of the respective binary nitrides, play an important role in determining the thermal stability. In particular, coatings with higher tantalum content consistently demonstrated enhanced nitrogen retention and structural integrity at elevated temperatures. This is likely due to tantalum's affinity for nitrogen and, even more importantly, its ability to stabilise nitride phases over a wide range of non-stoichiometry and specifically significant concentrations of nitrogen vacancies. 

Furthermore, it should be noted that entropic stabilisation is most effective under idealised conditions—e.g., perfect mixing, uniform distribution of atoms, and negligible stress effects, which are rarely achieved in sputtered thin films. Kinetic constraints, preferential bonding, surface energy effects, and residual stresses can all undermine the theoretical stabilisation expected from entropy alone. These findings underscore the importance of considering both enthalpic and entropic contributions when designing high-entropy thin films. 

\subsubsection{Mechanical properties}
The effective elastic modulus and hardness of all coatings that withstood at least the first annealing step are summarised in Table S1 and in Figure S9 of the Supplementary Material. Figure \ref{heat_mechprop} presents the results for coatings that endured all annealing steps.  
Annealing to 1000\textdegree C generally resulted in an increase in both effective elastic modulus and hardness relative to the as-deposited state, likely due to microstructural evolution and changes in internal stresses. Sample HT6 partially delaminated at this temperature, although it withstood all other annealing steps, probably due to a local defect in this particular specimen.  
Further annealing to 1200\textdegree C produced a minor decrease in both properties, consistent with the continued coarsening of the microstructure observed by SEM and supported by evidence of oxygen contamination. Stress relaxation may also contribute to this behaviour.  
Values measured after oxidation should be considered only indicative, as the coatings exhibited pronounced cracking, which is also reflected in the increased errors of the mechanical property measurements.

 \begin{figure}[htpb]
        \centering
        \includegraphics[width=.49\textwidth]{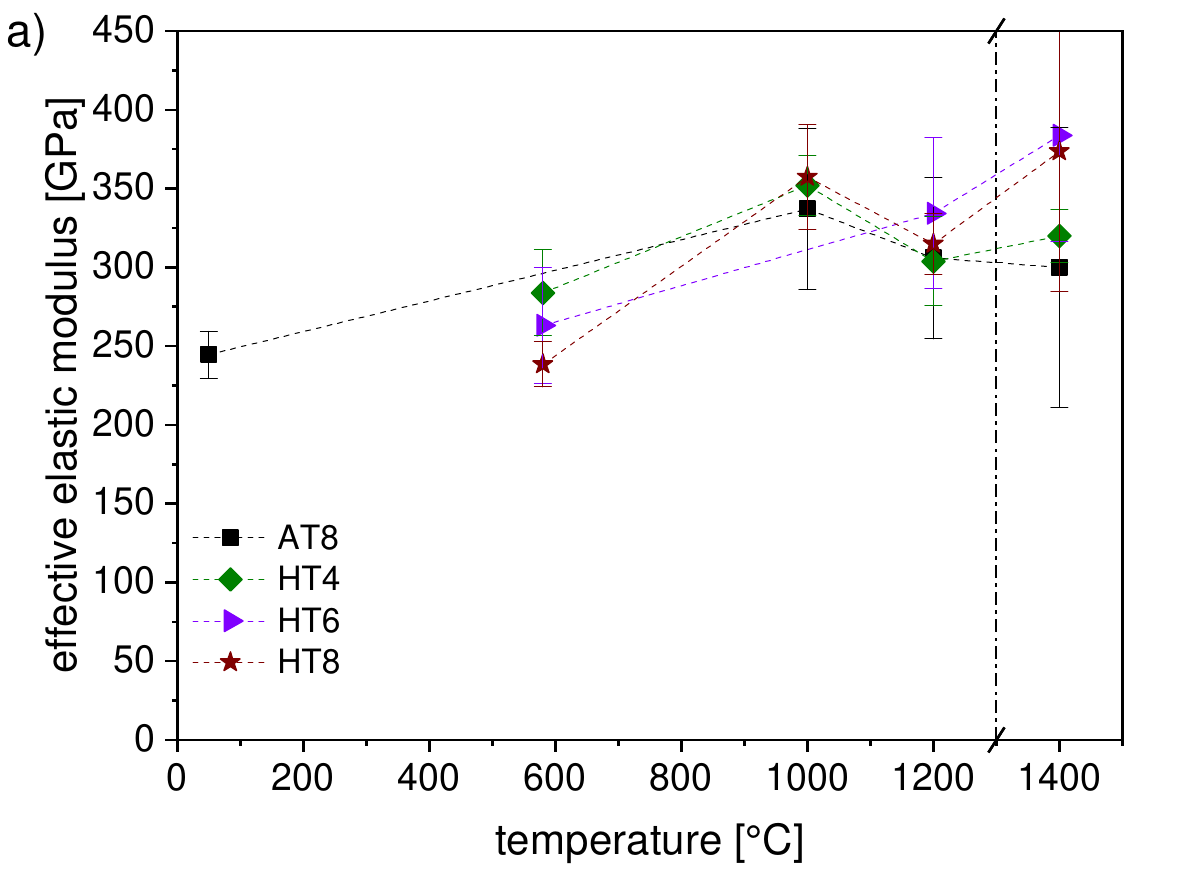}
        \includegraphics[width=.49\textwidth]{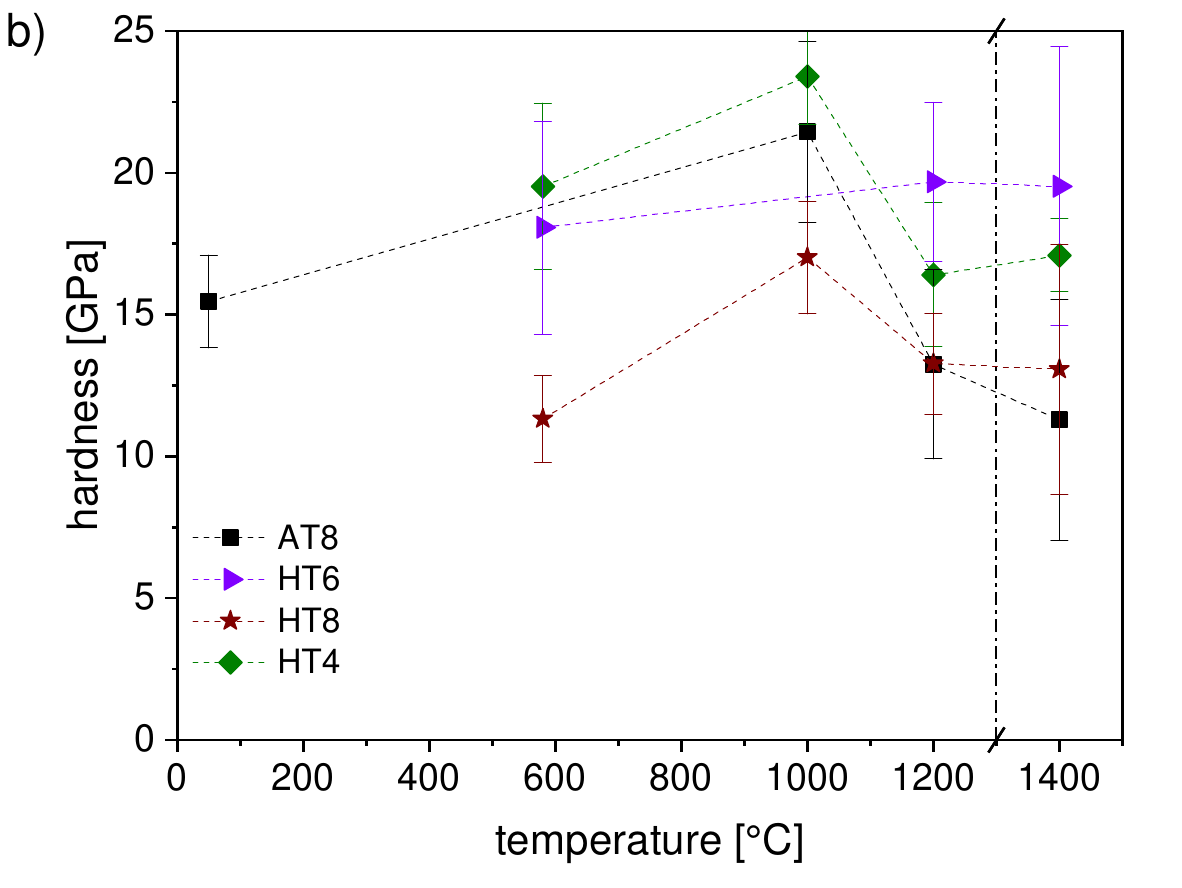}\\
        \caption{a) Effective elastic modulus and b) hardness of annealed coatings that withstood all annealing steps. Initial points represent the deposition temperature.}
        \label{heat_mechprop}
\end{figure}

\section{Conclusions}
Combining \textit{ab initio} and experimental techniques, high- and medium-entropy nitride coatings from the cubic (fcc) Cr–Hf–Mo–Ta–W–N systems were studied.
Calculations shed light on how individual elements affect phase stability and mechanical properties, suggesting a stabilising effect of nitrogen vacancies (in the range of $\approx{25}$\%), similar to some of the parent binary nitrides (TaN$_x$, MoN$_x$, WN$_x$). Hafnium and tantalum promoted chemical stability due to their strong nitrogen affinity, whereas tungsten---being unstable (mechanically and dynamically) in the fcc structure---increased formation energy, thus decreased stability the most pronouncedly. Molybdenum played neither a strongly stabilising nor a destabilising role.

The coatings were deposited by reactive DC magnetron sputtering from segmented elemental targets at an ambient temperature of $\sim$50\textdegree C (AT series) and at a high temperature of $\sim$580\textdegree C (HT series). The nitrogen content was generally below 50~at.\%, consistent with predictions that nitrogen vacancies stabilise the coatings. The HT series contained $\sim$5~at.\% less nitrogen than their AT counterparts, attributed to enhanced nitrogen adatom evaporation during high-temperature deposition.  

All coatings exhibited a columnar microstructure. The HT series showed dense columns, whereas the AT series displayed a looser structure with 1–2~nm wide amorphous column boundaries capable of adsorbing atmospheric oxygen, creating potential high-temperature failure sites. All coatings also showed compositional variation in the form of multilayering due to the experimental geometry.  

XRD analysis confirmed a highly textured single-phase fcc structure in all samples. One coating exhibited a (200) growth preference, consistent with surface-energy minimisation, while all others displayed strong (111) texture, where strain-energy minimisation is dominant. The measured lattice parameters (4.1–4.4~\AA) were governed by the atomic radii of the dominant elements and by the lattice parameters of the corresponding binary nitrides.  

Crystallite sizes in the AT series remained relatively stable at $\sim$40~nm, whereas high-temperature deposition promoted growth to $\sim$50–120~nm in the HT series. The denser structure of the HT coatings generally resulted in superior mechanical properties compared to the AT series, with hardness and elastic modulus trends correlating with the properties of the predominant elements’ binary nitrides.

The temperature stability of the coatings on silicon substrates was assessed by annealing at 1000\textdegree C and 1200\textdegree C. Oxidation was evaluated for coatings on sapphire substrates by annealing to 1400\textdegree C; after reaching the maximum temperature, the pressure was increased to 10\,Pa by introducing air for 3~minutes.  

All AT coatings, except for the medium-entropy AT8 (without chromium and hafnium), failed during the initial annealing step. The same occurred for the chromium- and hafnium-rich HT9 and the molybdenum-rich HT10, both prepared at high temperature. This behaviour highlights the importance of high-temperature deposition in producing a dense microstructure that enhances thermal stability. Among coatings that survived the first annealing step, no correlation was observed between thermal stability and mixing entropy. For example, while a nearly equimolar nitride failed at 1200\textdegree C, a medium-entropy coating with the lowest entropy readily withstood both the high-temperature anneal and the oxidation test.  

Annealing was characterised by morphological coarsening and nitrogen loss. Coatings with nitrogen depletion of $\gtrsim$20~at.\% after the first annealing step failed during the subsequent one, whereas those with losses of $\lesssim$10~at.\% survived all steps. The predominant elements proved critical for thermal stability. Hafnium, although calculated to be the strongest nitride former, remained stable only with a low concentration of nitrogen vacancies, limiting its ability to stabilise the coatings. In contrast, tantalum—despite being a weaker nitride former—can form stable nitrides even at very high nitrogen vacancy concentrations, and indeed, increased tantalum content was found to be essential for high-temperature stability. Chromium, by contrast, bonded only weakly with other elements and tended to diffuse out and evaporate during annealing.

While oxidation tests produced some changes in XRD patterns, the main features remained unaffected. Detailed structural modifications were revealed only by TEM. The dominant phase after oxidation was still an fcc nitride. Tungsten tended to segregate, forming an elemental metallic phase, while hafnium preferentially formed HfO$_2$ in multiple allotropes. These observations underscore the critical role of tantalum in achieving the best overall performance of the coatings.

\section*{CRediT authorship contribution statement}
\noindent Pavel Souček: Conceptualization, Formal analysis, Funding acquisition, Methodology, Project administration, Supervision, Visualization, Writing – original draft,  Writing - review \& editing.\\
Stanislava Debnárová: Formal analysis, Investigation, Visualization,  Writing - review \& editing.\\
Šárka zuzjaková: Conceptualization, Formal analysis, Investigation, Methodology, Writing – original draft,  Writing - review \& editing.\\
Shuyao Lin: Conceptualization, Formal analysis, Investigation, Methodology, Visualization, Writing – original draft,  Writing - review \& editing.\\
Matej Fekete: Formal analysis, Investigation,  Writing - review \& editing.\\
Zsolt Czigány: Conceptualization, Formal analysis, Investigation, Methodology, Visualization,  Writing - review \& editing.\\
Katalin Balázsi: Methodology, Supervision.\\
Lukáš Vrána: Formal analysis, Investigation, Visualization,  Writing - review \& editing.\\
Tatiana Pitoňáková: Formal analysis, Investigation, Visualization,  Writing - review \& editing.\\
Ondřej Jašek: Formal analysis, Investigation,  Writing - review \& editing.\\
Petr Zeman: Funding acquisition, Methodology, Supervision,  Writing - review \& editing.\\
Nikola Koutná: Conceptualization, Formal analysis, Investigation, Methodology, Supervision, Writing – original draft,  Writing - review \& editing.\\

\section*{Declaration of competing interest}
The authors declare that they have no known competing financial interests or personal relationships that could have appeared to influence the work reported in this paper.

\section*{Data availability}
Data will be made available on request.

\section*{Acknowledgment}
This research was supported by the project LM2023039 funded by the Ministry of Education, Youth, and Sports of the Czech Republic, project GA23-05947S financed by the Czech Science Foundation and by the grant no. VEKOP-2.3.3-15-2016-00002 and VEKOP-2.3.2-16-2016-00011 of the European Structural and Investment Funds. The research was also partly supported by the project QM4ST under Project No. CZ.02.01.01/00/22\_008/0004572 funded by Programme Johannes Amos Comenius, call Excellent Research.
The computations handling were enabled by resources provided by the National Academic Infrastructure for Supercomputing in Sweden (NAISS)---at supercomputer centers NSC and PDC, partially funded by the Swedish Research Council through grant agreement no. 2022-06725--and by the Vienna Scientific Cluster (VSC) in Austria. 
The authors acknowledge TU Wien Bibliothek for financial support through its Open Access Funding Program.



 \bibliographystyle{elsarticle-num} 
 \bibliography{cas-refs}

\begin{thebibliography}{10}
\expandafter\ifx\csname url\endcsname\relax
  \def\url#1{\texttt{#1}}\fi
\expandafter\ifx\csname urlprefix\endcsname\relax\def\urlprefix{URL }\fi
\expandafter\ifx\csname href\endcsname\relax
  \def\href#1#2{#2} \def\path#1{#1}\fi

\bibitem{CANTOR2004213}
B.~Cantor, I.~Chang, P.~Knight, A.~Vincent, Microstructural development in equiatomic multicomponent alloys, Mater. Sci. Eng. A 375-377 (2004) 213--218.
\newblock \href {https://doi.org/https://doi.org/10.1016/j.msea.2003.10.257} {\path{doi:https://doi.org/10.1016/j.msea.2003.10.257}}.

\bibitem{yeh04}
J.-W. Yeh, S.-K. Chen, S.-J. Lin, J.-Y. Gan, T.-S. Chin, T.-T. Shun, C.-H. Tsau, S.-Y. Chang, Nanostructured {H}igh-{E}ntropy {A}lloys with {M}ultiple {P}rincipal {E}lements: {N}ovel {A}lloy {D}esign {C}oncepts and {O}utcomes, Adv. Eng. Mater. 6~(5) (2004) 299--303.
\newblock \href {https://doi.org/https://doi.org/10.1002/adem.200300567} {\path{doi:https://doi.org/10.1002/adem.200300567}}.

\bibitem{MIRACLE2017448}
D.~Miracle, O.~Senkov, A critical review of high entropy alloys and related concepts, Acta Mater. 122 (2017) 448--511.
\newblock \href {https://doi.org/https://doi.org/10.1016/j.actamat.2016.08.081} {\path{doi:https://doi.org/10.1016/j.actamat.2016.08.081}}.

\bibitem{SENKOV2011698}
O.~Senkov, G.~Wilks, J.~Scott, D.~Miracle, Mechanical properties of {N}b$_{25}${M}o$_{25}${T}a$_{25}${W}$_{25}$ and {V}$_{20}${N}b$_{20}${M}o$_{20}${T}a$_{20}${W}$_{20}$ refractory high entropy alloys, Intermetallics 19~(5) (2011) 698--706.
\newblock \href {https://doi.org/https://doi.org/10.1016/j.intermet.2011.01.004} {\path{doi:https://doi.org/10.1016/j.intermet.2011.01.004}}.

\bibitem{STASIAK2022128987}
T.~Stasiak, P.~Souček, V.~Buršíková, N.~Koutná, Z.~Czigány, K.~Balázsi, P.~Vašina, Synthesis and characterization of the ceramic refractory metal high entropy nitride thin films from {C}r-{H}f-{M}o-{T}a-{W} system, Surf. Coat. Technol. 449 (2022) 128987.
\newblock \href {https://doi.org/https://doi.org/10.1016/j.surfcoat.2022.128987} {\path{doi:https://doi.org/10.1016/j.surfcoat.2022.128987}}.

\bibitem{oses2020high}
C.~Oses, C.~Toher, S.~Curtarolo, High-entropy ceramics, Nature Rev. Mater. 5~(4) (2020) 295--309.
\newblock \href {https://doi.org/https://doi.org/10.1038/s41578-019-0170-8} {\path{doi:https://doi.org/10.1038/s41578-019-0170-8}}.

\bibitem{VRANA2025131742}
L.~Vrána, T.~Stasiak, M.~Fekete, V.~Buršíková, Z.~Czigány, K.~Balázsi, P.~Souček, Effects of nitrogen content on microstructure and mechanical properties of {DC} magnetron sputtered {C}r-{M}n-{M}o-{S}i-{Y}-({N}) high entropy coatings, Surf. Coat. Technol. 497 (2025) 131742.
\newblock \href {https://doi.org/https://doi.org/10.1016/j.surfcoat.2025.131742} {\path{doi:https://doi.org/10.1016/j.surfcoat.2025.131742}}.

\bibitem{gromov2021structure}
V.~E. Gromov, S.~V. Konovalov, Y.~F. Ivanov, K.~Osintsev, Structure and properties of high-entropy alloys, Springer, 2021.
\newblock \href {https://doi.org/https://doi.org/10.1007/978-3-030-78364-8} {\path{doi:https://doi.org/10.1007/978-3-030-78364-8}}.

\bibitem{Braun2015}
A.~Y.~C. Nee (Ed.), Handbook of Manufacturing Engineering and Technology, Springer London, London, 2015.
\newblock \href {https://doi.org/https://doi.org/10.1007/978-1-4471-4670-4\_28} {\path{doi:https://doi.org/10.1007/978-1-4471-4670-4\_28}}.

\bibitem{HE2021}
C.-Y. He, X.-H. Gao, D.-M. Yu, S.-S. Zhao, H.-X. Guo, G.~Liu, Toward high-temperature thermal tolerance in solar selective absorber coatings: choosing high entropy ceramic {H}f{N}b{T}a{T}i{Z}r{N}, J. Mater. Chem. A 9 (2021) 21270--21280.
\newblock \href {https://doi.org/10.1039/D1TA06682J} {\path{doi:10.1039/D1TA06682J}}.

\bibitem{GRUBER2023130016}
G.~C. Gruber, S.~Wurster, M.~J. Cordill, R.~Franz, Refractory high entropy metal sublattice nitride thin films as diffusion barriers in {C}u metallizations, Surf. Coat. Technol. 473 (2023) 130016.
\newblock \href {https://doi.org/https://doi.org/10.1016/j.surfcoat.2023.130016} {\path{doi:https://doi.org/10.1016/j.surfcoat.2023.130016}}.

\bibitem{Scherrer1918}
P.~Scherrer, Bestimmung der {G}röße und der inneren {S}truktur von {K}olloidteilchen mittels {R}öntgenstrahlen, G{\"o}tt. Nachr. 1918 (1918) 98--100.

\bibitem{oliver1992improved}
W.~C. Oliver, G.~M. Pharr, An improved technique for determining hardness and elastic modulus using load and displacement sensing indentation experiments, Journal of materials research 7~(6) (1992) 1564--1583.
\newblock \href {https://doi.org/https://doi.org/10.1557/JMR.1992.1564} {\path{doi:https://doi.org/10.1557/JMR.1992.1564}}.

\bibitem{labar2012electron}
J.~L. L{\'a}b{\'a}r, M.~Adamik, B.~Barna, Z.~Czig{\'a}ny, Z.~Fogarassy, Z.~E. Horv{\'a}th, O.~Geszti, F.~Misj{\'a}k, J.~Morgiel, G.~Radn{\'o}czi, et~al., Electron diffraction based analysis of phase fractions and texture in nanocrystalline thin films, part {III: A}pplication examples, Microsc. Microanal. 18~(2) (2012) 406--420.
\newblock \href {https://doi.org/https://doi.org/10.1017/S1431927611012803} {\path{doi:https://doi.org/10.1017/S1431927611012803}}.

\bibitem{VASP-1}
G.~Kresse, J.~Furthm{\"u}ller, Efficient iterative schemes for ab initio total-energy calculations using a plane-wave basis set, Phys. Rev. B 54~(16) (1996) 11169.

\bibitem{VASP-2}
G.~Kresse, D.~Joubert, From ultrasoft pseudopotentials to the projector augmented-wave method, Phys. Rev. B 59 (1999) 1758--1775.

\bibitem{PBEsol}
J.~P. Perdew, A.~Ruzsinszky, G.~I. Csonka, O.~A. Vydrov, G.~E. Scuseria, L.~A. Constantin, X.~Zhou, K.~Burke, Restoring the density-gradient expansion for exchange in solids and surfaces, Phys. Rev. Lett. 100 (2008) 136406.

\bibitem{stasiak2024synthesis}
T.~Stasiak, S.~Debn{\'a}rov{\'a}, S.~Lin, N.~Koutn{\'a}, Z.~Czig{\'a}ny, K.~Bal{\'a}zsi, V.~Bur{\v{s}}{\'\i}kov{\'a}, P.~Va{\v{s}}ina, P.~Sou{\v{c}}ek, Synthesis and characterization of ceramic high entropy carbide thin films from the cr-hf-mo-ta-w refractory metal system, Surface and Coatings Technology 485 (2024) 130839.

\bibitem{SQS-1}
S.-H. Wei, L.~Ferreira, J.~E. Bernard, A.~Zunger, Electronic properties of random alloys: {S}pecial quasirandom structures, Phys. Rev. B 42~(15) (1990) 9622.

\bibitem{SQS-2}
A.~Zunger, S.-H. Wei, L.~Ferreira, J.~Bernard, Special quasirandom structures, Phys. Rev. Lett. 65~(3) (1990) 353.

\bibitem{koutna2024phase}
N.~Koutn{\'a}, L.~Hultman, P.~H. Mayrhofer, D.~G. Sangiovanni, Phase stability and mechanical property trends for mab phases by high-throughput ab initio calculations, Materials \& Design 241 (2024) 112959.

\bibitem{le2001symmetry}
Y.~Le~Page, P.~Saxe, Symmetry-general least-squares extraction of elastic coefficients from ab initio total energy calculations, Physical Review B 63~(17) (2001) 174103.

\bibitem{koutna2021high}
N.~Koutn{\'a}, A.~Brenner, D.~Holec, P.~H. Mayrhofer, High-throughput first-principles search for ceramic superlattices with improved ductility and fracture resistance, Acta materialia 206 (2021) 116615.

\bibitem{nye1985physical}
J.~F. Nye, Physical properties of crystals: their representation by tensors and matrices, Oxford university press, 1985.

\bibitem{mouhat2014necessary}
F.~Mouhat, F.-X. Coudert, Necessary and sufficient elastic stability conditions in various crystal systems, Phys. Rev. B 90~(22) (2014) 224104.

\bibitem{koutna2016point}
N.~Koutn{\'a}, D.~Holec, O.~Svoboda, F.~F. Klimashin, P.~H. Mayrhofer, Point defects stabilise cubic {Mo}--{N} and {Ta}--{N}, Journal of Physics D: Applied Physics 49~(37) (2016) 375303.

\bibitem{balasubramanian2016vacancy}
K.~Balasubramanian, S.~Khare, D.~Gall, Vacancy-induced mechanical stabilization of cubic tungsten nitride, Physical Review B 94~(17) (2016) 174111.

\bibitem{lee2021defect}
G.~Lee, H.~Lee, T.~Lee, A.~Soon, Defect-mediated ab initio thermodynamics of metastable $\gamma$-{MoN} (001), The Journal of Chemical Physics 154~(6) (2021).

\bibitem{klimashin2016impact}
F.~F. Klimashin, N.~Koutn{\'a}, H.~Euchner, D.~Holec, P.~H. Mayrhofer, The impact of nitrogen content and vacancies on structure and mechanical properties of {Mo}--{N} thin films, Journal of Applied Physics 120~(18) (2016).

\bibitem{ozsdolay2017cation}
B.~Ozsdolay, K.~Balasubramanian, D.~Gall, Cation and anion vacancies in cubic molybdenum nitride, Journal of Alloys and Compounds 705 (2017) 631--637.

\bibitem{ozsdolay2016cubic}
B.~Ozsdolay, C.~Mulligan, K.~Balasubramanian, L.~Huang, S.~Khare, D.~Gall, Cubic $\beta$-{WN}$_x$ layers: {G}rowth and properties vs {N}-to-{W} ratio, Surface and Coatings Technology 304 (2016) 98--107.

\bibitem{buchinger2019toughness}
J.~Buchinger, N.~Koutn{\'a}, Z.~Chen, Z.~Zhang, P.~H. Mayrhofer, D.~Holec, M.~Bartosik, Toughness enhancement in {TiN}/{WN} superlattice thin films, Acta Materialia 172 (2019) 18--29.

\bibitem{zhang2013insights}
Z.~Zhang, H.~Li, R.~Daniel, C.~Mitterer, G.~Dehm, Insights into the atomic and electronic structure triggered by ordered nitrogen vacancies in {CrN}, Physical Review B—Condensed Matter and Materials Physics 87~(1) (2013) 014104.

\bibitem{SAFI2000203}
I.~Safi, Recent aspects concerning {DC} reactive magnetron sputtering of thin films: a review, Surf. Coat. Technol. 127~(2) (2000) 203--218.
\newblock \href {https://doi.org/https://doi.org/10.1016/S0257-8972(00)00566-1} {\path{doi:https://doi.org/10.1016/S0257-8972(00)00566-1}}.

\bibitem{LIANG20117709}
S.-C. Liang, Z.-C. Chang, D.-C. Tsai, Y.-C. Lin, H.-S. Sung, M.-J. Deng, F.-S. Shieu, Effects of substrate temperature on the structure and mechanical properties of ({T}i{VC}r{Z}r{H}f){N} coatings, Appl. Surf. Sci. 257~(17) (2011) 7709--7713.
\newblock \href {https://doi.org/https://doi.org/10.1016/j.apsusc.2011.04.014} {\path{doi:https://doi.org/10.1016/j.apsusc.2011.04.014}}.

\bibitem{VONFIEANDT2020137685}
K.~{von Fieandt}, E.-M. Paschalidou, A.~Srinath, P.~Soucek, L.~Riekehr, L.~Nyholm, E.~Lewin, Multi-component ({A}l,{C}r,{N}b,{Y},{Z}r){N} thin films by reactive magnetron sputter deposition for increased hardness and corrosion resistance, Thin Solid Films 693 (2020) 137685.
\newblock \href {https://doi.org/https://doi.org/10.1016/j.tsf.2019.137685} {\path{doi:https://doi.org/10.1016/j.tsf.2019.137685}}.

\bibitem{ANDERS20104087}
A.~Anders, A structure zone diagram including plasma-based deposition and ion etching, Thin Solid Films 518~(15) (2010) 4087--4090.
\newblock \href {https://doi.org/https://doi.org/10.1016/j.tsf.2009.10.145} {\path{doi:https://doi.org/10.1016/j.tsf.2009.10.145}}.

\bibitem{LAI20063275}
C.-H. Lai, S.-J. Lin, J.-W. Yeh, S.-Y. Chang, Preparation and characterization of {A}l{C}r{T}a{T}i{Z}r multi-element nitride coatings, Surf. Coat. Technol. 201~(6) (2006) 3275--3280.
\newblock \href {https://doi.org/https://doi.org/10.1016/j.surfcoat.2006.06.048} {\path{doi:https://doi.org/10.1016/j.surfcoat.2006.06.048}}.

\bibitem{REN2013171}
B.~Ren, Z.~Shen, Z.~Liu, Structure and mechanical properties of multi-element ({A}l{C}r{M}n{M}o{N}i{Z}r){N}$_x$ coatings by reactive magnetron sputtering, J. Alloy. Comp. 560 (2013) 171--176.
\newblock \href {https://doi.org/https://doi.org/10.1016/j.jallcom.2013.01.148} {\path{doi:https://doi.org/10.1016/j.jallcom.2013.01.148}}.

\bibitem{PELLEG1991117}
J.~Pelleg, L.~Zevin, S.~Lungo, N.~Croitoru, Reactive-sputter-deposited {T}i{N} films on glass substrates, Thin Solid Films 197~(1) (1991) 117--128.
\newblock \href {https://doi.org/https://doi.org/10.1016/0040-6090(91)90225-M} {\path{doi:https://doi.org/10.1016/0040-6090(91)90225-M}}.

\bibitem{matproj}
A.~Jain, S.~P. Ong, G.~Hautier, W.~Chen, W.~D. Richards, S.~Dacek, S.~Cholia, D.~Gunter, D.~Skinner, G.~Ceder, K.~A. Persson, Commentary: {T}he {M}aterials {P}roject: {A} materials genome approach to accelerating materials innovation, APL Mater. 1~(1) (2013) 011002.
\newblock \href {https://doi.org/10.1063/1.4812323} {\path{doi:10.1063/1.4812323}}.

\bibitem{KUNC2003744}
F.~Kunc, J.~Musil, P.~Mayrhofer, C.~Mitterer, Low-stress superhard {T}i{B} films prepared by magnetron sputtering, Surface and Coatings Technology 174-175 (2003) 744--753, proceedings of the Eight International Conference on Plasma Surface Engineering.
\newblock \href {https://doi.org/https://doi.org/10.1016/S0257-8972(03)00425-0} {\path{doi:https://doi.org/10.1016/S0257-8972(03)00425-0}}.

\bibitem{HERAULT2021117385}
Q.~Hérault, I.~Gozhyk, M.~Balestrieri, H.~Montigaud, S.~Grachev, R.~Lazzari, Kinetics and mechanisms of stress relaxation in sputtered silver thin films, Acta Materialia 221 (2021) 117385.
\newblock \href {https://doi.org/https://doi.org/10.1016/j.actamat.2021.117385} {\path{doi:https://doi.org/10.1016/j.actamat.2021.117385}}.

\bibitem{birkholz2006thin}
M.~Birkholz, Thin Film Analysis by X-Ray Scattering, Vol. 252, Wiley-VCH Verlag GmbH \& Co. KGaA, 2006.

\bibitem{sarmast2023residual}
A.~Sarmast, J.~Schubnell, J.~Preu{\ss}ner, M.~Hinterstein, E.~Carl, Residual stress analysis in industrial parts: A comprehensive comparison of xrd methods, J. Mater. Sci. 58~(44) (2023) 16905--16929.
\newblock \href {https://doi.org/https://doi.org/10.1007/s10853-023-09069-z} {\path{doi:https://doi.org/10.1007/s10853-023-09069-z}}.

\bibitem{Levi_1997}
A.~C. Levi, M.~Kotrla, Theory and simulation of crystal growth, J. Phys.-Condens. Mat. 9~(2) (1997) 299.
\newblock \href {https://doi.org/https://doi.org/10.1088/0953-8984/9/2/001} {\path{doi:https://doi.org/10.1088/0953-8984/9/2/001}}.

\bibitem{KINDLUND2019137479}
H.~Kindlund, D.~Sangiovanni, I.~Petrov, J.~Greene, L.~Hultman, A review of the intrinsic ductility and toughness of hard transition-metal nitride alloy thin films, Thin Solid Films 688 (2019) 137479, a Special Issue “Thin Film Advances”, dedicated to the 75th birthday of Professor Joe Greene.
\newblock \href {https://doi.org/https://doi.org/10.1016/j.tsf.2019.137479} {\path{doi:https://doi.org/10.1016/j.tsf.2019.137479}}.

\bibitem{hones2003structural}
P.~Hones, N.~Martin, M.~Regula, F.~L{\'e}vy, Structural and mechanical properties of chromium nitride, molybdenum nitride, and tungsten nitride thin films, J. Phys. D 36~(8) (2003) 1023.
\newblock \href {https://doi.org/https://doi.org/10.1088/0022-3727/36/8/313} {\path{doi:https://doi.org/10.1088/0022-3727/36/8/313}}.

\bibitem{HU2015141}
C.~Hu, X.~Zhang, Z.~Gu, H.~Huang, S.~Zhang, X.~Fan, W.~Zhang, Q.~Wei, W.~Zheng, Negative effect of vacancies on cubic symmetry, hardness and conductivity in hafnium nitride films, Scripta Mater. 108 (2015) 141--146.
\newblock \href {https://doi.org/https://doi.org/10.1016/j.scriptamat.2015.07.002} {\path{doi:https://doi.org/10.1016/j.scriptamat.2015.07.002}}.

\bibitem{BERNOULLI2013157}
D.~Bernoulli, U.~Müller, M.~Schwarzenberger, R.~Hauert, R.~Spolenak, Magnetron sputter deposited tantalum and tantalum nitride thin films: An analysis of phase, hardness and composition, Thin Solid Films 548 (2013) 157--161.
\newblock \href {https://doi.org/https://doi.org/10.1016/j.tsf.2013.09.055} {\path{doi:https://doi.org/10.1016/j.tsf.2013.09.055}}.

\bibitem{hall1951deformation}
E.~Hall, The deformation and ageing of mild steel: {III} discussion of results, Proc. Phys. Soc. B 64~(9) (1951) 747.
\newblock \href {https://doi.org/https://10.1088/0370-1301/64/9/303} {\path{doi:https://10.1088/0370-1301/64/9/303}}.

\bibitem{MA2004184}
C.-H. Ma, J.-H. Huang, H.~Chen, Texture evolution of transition-metal nitride thin films by ion beam assisted deposition, Thin Solid Films 446~(2) (2004) 184--193.
\newblock \href {https://doi.org/https://doi.org/10.1016/j.tsf.2003.09.063} {\path{doi:https://doi.org/10.1016/j.tsf.2003.09.063}}.

\bibitem{TOROK198737}
E.~Török, A.~Perry, L.~Chollet, W.~Sproul, Young's modulus of {T}i{N}, {T}i{C}, {Z}r{N} and {H}f{N}, Thin Solid Films 153~(1) (1987) 37--43.
\newblock \href {https://doi.org/https://doi.org/10.1016/0040-6090(87)90167-2} {\path{doi:https://doi.org/10.1016/0040-6090(87)90167-2}}.

\bibitem{dastan2022influence}
D.~Dastan, K.~Shan, A.~Jafari, F.~Gity, X.-T. Yin, Z.~Shi, N.~D. Alharbi, B.~A. Reshi, W.~Fu, {\c{S}}.~{\c{T}}{\u{a}}lu, et~al., Influence of nitrogen concentration on electrical, mechanical, and structural properties of tantalum nitride thin films prepared via {DC} magnetron sputtering, Appl. Phys. A 128~(5) (2022) 400.
\newblock \href {https://doi.org/https://doi.org/10.1007/s00339-022-05501-4} {\path{doi:https://doi.org/10.1007/s00339-022-05501-4}}.

\bibitem{SUN2024130775}
S.~Sun, H.~Wang, L.~Huang, Z.~Feng, R.~Sun, W.~Zhang, W.~Zhang, Microstructure evolution and mechanical properties of refractory high-entropy alloy nitride film, Surf. Coat. Technol. 483 (2024) 130775.
\newblock \href {https://doi.org/https://doi.org/10.1016/j.surfcoat.2024.130775} {\path{doi:https://doi.org/10.1016/j.surfcoat.2024.130775}}.

\bibitem{sheng2016nano}
W.~Sheng, X.~Yang, C.~Wang, Y.~Zhang, Nano-crystallization of high-entropy amorphous {N}b{T}i{A}l{S}i{W}$_x${N}$_y$ films prepared by magnetron sputtering, Entropy 18~(6) (2016) 226.
\newblock \href {https://doi.org/https://doi.org/10.3390/e18060226} {\path{doi:https://doi.org/10.3390/e18060226}}.

\bibitem{Pettifor01041992}
D.~G. Pettifor, Theoretical predictions of structure and related properties of intermetallics, Mater. Sci. Technol. 8~(4) (1992) 345--349.
\newblock \href {https://doi.org/10.1179/mst.1992.8.4.345} {\path{doi:10.1179/mst.1992.8.4.345}}.

\bibitem{LEYLAND20001}
A.~Leyland, A.~Matthews, On the significance of the {H}/{E} ratio in wear control: a nanocomposite coating approach to optimised tribological behaviour, Wear 246~(1) (2000) 1--11.
\newblock \href {https://doi.org/https://doi.org/10.1016/S0043-1648(00)00488-9} {\path{doi:https://doi.org/10.1016/S0043-1648(00)00488-9}}.

\bibitem{musil}
J.~Musil, Flexible hard nanocomposite coatings, RSC Adv. 5 (2015) 60482--60495.
\newblock \href {https://doi.org/https://doi.org/10.1039/C5RA09586G} {\path{doi:https://doi.org/10.1039/C5RA09586G}}.

\bibitem{MAYRHOFER2003725}
P.~Mayrhofer, C.~Mitterer, J.~Musil, Structure–property relationships in single- and dual-phase nanocrystalline hard coatings, Surf. Coat. Technol. 174-175 (2003) 725--731.
\newblock \href {https://doi.org/https://doi.org/10.1016/S0257-8972(03)00576-0} {\path{doi:https://doi.org/10.1016/S0257-8972(03)00576-0}}.

\bibitem{MURTY1998328}
M.~Murty, B.~Cowles, B.~Cooper, Surface smoothing during sputtering: mobile vacancies versus adatom detachment and diffusion, Surf. Sci. 415~(3) (1998) 328--335.
\newblock \href {https://doi.org/https://doi.org/10.1016/S0039-6028(98)00556-1} {\path{doi:https://doi.org/10.1016/S0039-6028(98)00556-1}}.

\bibitem{eltoukhy79}
A.~H. Eltoukhy, J.~E. Greene, Compositionally modulated sputtered {I}n{S}b/{G}a{S}b superlattices: {C}rystal growth and interlayer diffusion, J. Appl. Phys. 50~(1) (1979) 505--517.
\newblock \href {https://doi.org/https://doi.org/10.1063/1.325643} {\path{doi:https://doi.org/10.1063/1.325643}}.

\bibitem{Windischmann01011992}
H.~Windischmann, Intrinsic stress in sputter-deposited thin films, Crit. Rev. Solid State 17~(6) (1992) 547--596.
\newblock \href {https://doi.org/https://doi.org/10.1080/10408439208244586} {\path{doi:https://doi.org/10.1080/10408439208244586}}.

\bibitem{WARRES2023139977}
C.~Warres, J.~Meyer, T.~Lutz, P.~Albrecht, B.~Schröppel, W.~Engelhart, J.~Kümmel, Ion- and temperature-induced 3-dimensional nanoscale patterning in {T}i$_{1-x}${A}l$_x${N} deposited by {H}igh {P}ower {I}mpulse {M}agnetron {S}puttering, Thin Solid Films 781 (2023) 139977.
\newblock \href {https://doi.org/https://doi.org/10.1016/j.tsf.2023.139977} {\path{doi:https://doi.org/10.1016/j.tsf.2023.139977}}.

\bibitem{Widenmeyer2014}
M.~Widenmeyer, E.~Meissner, A.~Senyshyn, R.~Niewa, On the formation mechanism of chromium nitrides: An in situ study, Z. Anorg. Allg. Chem. 640~(14) (2014) 2801--2808.
\newblock \href {https://doi.org/https://doi.org/10.1002/zaac.201400246} {\path{doi:https://doi.org/10.1002/zaac.201400246}}.

\end{thebibliography}





\end{document}